\documentclass[pra, aps, twocolumn, nofootinbib, tightenlines, nobibnotes, showpacs, 10pt]{revtex4-1}

\usepackage{bigints}
\usepackage{amssymb}
\usepackage{graphicx}
\usepackage{amsmath}
\usepackage{dcolumn}
\usepackage{bm}
\usepackage{multirow}
\usepackage{hyperref}
\usepackage{physics}
\usepackage{bm}
\usepackage{comment}
\usepackage{xcolor}
\usepackage[caption=false]{subfig}
\usepackage{float}
\usepackage{siunitx}
\usepackage{dsfont}
\usepackage{bbold}
\usepackage{appendix}
\usepackage{relsize}
\usepackage{leftidx}

\definecolor{YKblue}{rgb}{0.0, 0.18, 0.65}
\definecolor{carmine}{rgb}{0.81, 0.09, 0.13}
\definecolor{lavender}{rgb}{0.48, 0.03, 0.89}

\hypersetup{
colorlinks=true,
linkcolor=YKblue,
linktoc=page,
citecolor=YKblue,
urlcolor=YKblue
}

\everymath{\displaystyle}

\newcommand{\ov}[1]{\overline{#1}}

\usepackage{tikz}
\usetikzlibrary{patterns,arrows,shapes,arrows,intersections,decorations}

\begin{document}
\title{Wigner--Weyl description of massless Dirac plasmas}
	\author{Jos\'e L. Figueiredo}
	\email{jose.luis.figueiredo@tecnico.ulisboa.pt} 
		
	\author{Jo\~{a}o P. S. Bizarro}
 \email{bizarro@ipfn.tecnico.ulisboa.pt}
	
	\author{Hugo Ter\c{c}as}
	\email{hugo.tercas@tecnico.ulisboa.pt}
	
	\affiliation{Instituto de Plasmas e Fus\~{a}o Nuclear, Instituto
	Superior T\'{e}cnico, Universidade de Lisboa, 1049-001 Lisboa, Portugal}

\begin{abstract}

We derive a quantum kinetic model describing the dynamics of graphene electrons in phase space based on the Wigner--Weyl formalism. To take into account the quantum nature of the carriers, we make use of the quantum Liouville equation for the density matrix. By relating the density matrix elements with the Wigner function, the equation of motion for the latter is established, with the Coulomb interaction being introduced self-consistently (i.e., in the Hartree approximation). The long-wavelength limit for the plasmon dispersion relation is obtained, for both ungated and gated situations. As an application, we derive the corresponding fluid equations from first principles and discuss the correct value of the effective hydrodynamic mass of the carriers. This constitutes a crucial point in establishing the appropriate fluid description of Dirac electrons, thus paving the way to a more comprehensive description of graphene plasmonics. 	
\end{abstract}	

\maketitle

\section{Introduction}

In recent years, graphene has been extensively studied due to its outstanding optical, electronic and mechanical properties \cite{Castroetall}. In addition to its two-dimensional (2D) nature, the elementary electronic excitations are described by a Dirac-like dispersion in the low-energy limit \cite{Wallace}. The relativistic nature of graphene electrons, resulting from the cone-like dispersion relation near the Dirac points, makes it useful for transparent electronic devices, ultra-sensitive photodetectors, and other high-performance optoelectronic structures \cite{sensing1,sensing2,sensing3}. Furthermore, graphene possesses extremely high quantum efficiency for light-matter interactions \cite{Koppens}. Also, the collective oscillations of the electron and hole densities lead to the formation of plasmons (or plasma waves) \cite{plasmonics1}. In fact, plasmonic of 2D materials, such as graphene, transition metal dichalcogenides (TMDCs), and hexagonal boron nitride (hBN) \cite{Dai1125, Lin2009}, is nowadays a very prominent field of research \cite{plasmonics2}. Graphene-based plasmonics finds a variety of applications, as the versatility of graphene enables the manufacture of optical devices working in different frequency ranges, namely in the terahertz (THz) and the infra-red domains \cite{Grigorenko2012}. While metal plasmonics exhibit large Ohmic losses, which limits their applicability to optical processing devices, doped graphene emerges as an alternative. Its large conductivity, in part due to the zero-mass character of the carriers, encloses a wide range of potential applications, such as high-frequency nanoelectronics, nanomechanics, transparent electrodes, and composite materials \cite{Geim1}. For this reason, the possibility of electric gating has been extensively studied in graphene, allowing for the manipulation of the Fermi level \cite{gating}. Recently, gating with a solid electrolyte allowed carrier concentrations as large as $10^{14} \ \text{cm$^{-2}$}$ to be achieved, which results in a Fermi energy of $\mathcal E_\text{F} \simeq 1$ eV, such that a modulation of optical transmission in the visible spectrum is possible \cite{Wang,lawrence}. The potentiality for THz emission, by making use of a graphene field-effect transistor (gFET) and by controlling the applied gate voltage and injected current, has also been recently pointed out as a possible application \cite{graf4}.

From the theoretical point of view, a variety of techniques have been developed to establish the dynamics of Dirac fermions in graphene, ranging from semiclassical hydrodynamical models \cite{graf1,graf2,graf3} to quantum formulations that involve collective Green's functions, such as the time-dependent Hartree-Fock approximation \cite{HFintroduction1,HFintroduction2}, or the time-dependent density functional theory \cite{DFTintroduction}. Whilst it is often the case that cumbersome equations, of very reduced utility, crop up when going for a complete quantum description, it is also true that the semiclassical approach, based on the Vlasov equation, is inadequate in the low temperature or high density regimes \cite{dugaev_2013}. 

In this work, we establish a kinetic formalism based on the Wigner--Weyl formulation of quantum mechanics to study a Dirac plasma in phase space \cite{wigner}. We start by setting an equation of motion for the density matrix in the conduction and valence basis, derived from a microscopic tight-binding model for the low-energy electrons. Then, we derive a kinetic equation for the Wigner function components, which are related with the density matrix and incorporate the pseudo-spin degrees of freedom. Moreover, the interaction is introduced self-consistently via the Hartree approximation, which obeys the Poisson equation. Note that, despite previous works have already been focused on the Wigner representation for the case of graphene \cite{ans1,ans2}, our approach manages to treat the potential term as a perturbation, by making use of the conduction and valence eigen-states as the natural basis. By following this strategy, we are able to obtain the plasmon dispersion relation, recovering the usual result based on the random-phase approximation (RPA). By performing averages over the phase-space distributions (more precisely, by taking the moments of the evolution equation for the Wigner equation), hydrodynamical equations are obtained allowing for a fluid description of the Dirac particles in graphene. As a consequence, we are able to derive microscopically the effective mass of a fluid particle and relate it to Drude's mass, thus contributing to the understanding of a still-open question in graphene hydrodynamics. We show as well that the classical limit appropriately recovers results that have been previously obtained based on the Vlasov equation for the classical distribution function. 

So, the paper is organised as follows: in Sec. \ref{sec_preliminaries}, we review the basic properties of graphene carriers and corresponding low-energy Hamiltonian. In Sec. \ref{sec_wigner}, we derive the quantum kinetic equation based on the Wigner--Weyl approach for the Dirac plasma. In Sec. \ref{sec_hydrodynamics}, we obtain the fluid equations, valid at the macroscopic scale, by taking the moments from the previously obtained kinetic equation, and discriminate its classical and quantum contributions. As a consequence, we put forward an expression for the hydrodynamical mass relating the fluid momentum and velocity fields. As an example, we correct the plasmon dispersion relation and show the emergence of a purely quantum contribution, not revealed by RPA calculations. Finally, in Sec. \ref{sec_conclusions}, some conclusions are drawn and future perspectives are outlined. 

\section{Graphene preliminaries}
\label{sec_preliminaries} 

The electronic dynamics can be captured starting with a general form for the free Hamiltonian, written in a second quantized fashion as
\begin{equation}
\widehat{H}_0 = \sum_{ll'} \sum_{\bm{R}\bm{R'}} \widehat{u}_l^{\dag}(\bm{R}) \matrixel{\widehat u_l,\bm{R}}{\widehat{H}}{\widehat u_{l'},\bm{R'}}\widehat u_{l'}(\bm{R'}) ,
\label{TBhamiltonian}
\end{equation}
where $\bm{R}$ and $\bm{R'}$ run over the graphene lattice, and $l$ and $l'$ over the two sublattices ($A$ and $B$). Additionally, $\widehat u_l^\dag(\bm R)$ and $\widehat u_l(\bm R)$ denote the creation and annihilation operators, respectively, for each lattice point $\bm R$ and sublattice $l$. Moreover, the tight-binding approximation can be settled with a proper restriction on the matrix elements $\matrixel{\widehat u_l,\bm{R}}{\widehat{H}}{\widehat u_{l'},\bm{R'}}$. By allowing hopping only between nearest neighbors, we can set all matrix elements to zero with the exception of the cases of $\matrixel{\widehat u_l,\bm{R}}{\widehat{H}}{\widehat u_{l'},\bm{R}+ \bm{\delta}_i} = -t(1-\delta_{ll'})$, with $t\simeq 2.97 \text{ eV}$ the hopping integral and $\bm{\delta}_i$ the nearest-neighbor vectors \cite{Wallace}. Resorting to the relation $\widehat{u}_l(\bm R)= \sum_{\bm k} \widehat{u}_{l\bm k}e^{i\bm k \cdot \bm R}/\sqrt{N}$, where $N$ is the total number of carbon atoms, Eq.~\eqref{TBhamiltonian} reduces to
\begin{equation}
\widehat{H}_0 = \sum_{\bm{k}} \ \widehat{\bm{\varphi}}_{\bm{k}}^{\dag} \ \left(\begin{array}{cc} 0 & -t\Delta\\ -t\Delta^{\ast} &0\end{array}\right) \ \widehat{\bm{\varphi}}_{\bm{k}},
\label{ABhamiltonian}
\end{equation}
where $\widehat{\bm{\varphi}}_{\bm{k}} = (\widehat u_{A\bm{k}},\widehat u_{B\bm{k}})^{T}$ and $\Delta = \sum_{i}e^{-i\bm{k}\cdot\bm{\delta}_i}$. By working in the single-particle approximation, we can concentrate on a fixed wave-vector in the above summation. As such, the effective free Hamiltonian that we adopt comes after expanding Eq.~\eqref{ABhamiltonian} around the Dirac point $\bm K=(4\pi/3\sqrt{3}d,0)$, keeping only the linear terms and project it onto a fixed $\bm k$, obtaining \begin{equation}
	\widehat{H}_0 = \hbar v_{\rm{F}} \bm \sigma \cdot \widehat{\bm k} = \hbar v_{\rm{F}} \left(\begin{array}{cc} 
			0 & \widehat{k}_x - i \widehat{k}_y \\
			\widehat{k}_x + i\widehat{k}_y & 0
		\end{array} \right) ,\label{CVhamiltonian}
\end{equation}
with $\widehat{\bm k} = (\widehat{k}_x,\widehat{k}_y)$ the 2D wave-vector operator, $\bm \sigma=(\sigma_x,\sigma_y)$ the 2D vector of Pauli matrices, $v_\text{F} = 3ta/(2\hbar) \simeq 10^6 \text{ ms$^{-1}$}$ the Fermi velocity, and $a\simeq 2.6\text{ \si{\angstrom}}$ the lattice parameter. It is important to note that Eq.~\eqref{CVhamiltonian} is appropriate to study the transport properties in graphene, specially in the semiclassical limit, but does not capture the full collective behaviour. However, for the purposes of the current work, it suffices to remain within this level of approximation. After diagonalizing Eq.~\eqref{CVhamiltonian}, we find its two eigen-pseudo-spinors
\begin{equation}
	\ket{s}_{\bm k} = \frac{1}{\sqrt 2} \left(\begin{array}{c} 1 \\ s e^{i\phi_{\bm k}} \end{array}\right),
\end{equation}
where $s=\pm 1$ labels the conduction $(+1)$ and valence $(-1)$ bands and $\phi_{\bm k} = \arctan(k_y/k_x)$. We have $\widehat{H}_0 \ket{s}_{\bm k} = \mathcal E_{\bm k}^s \ket{s}_{\bm k}$, where $\mathcal E_{\bm k}^s = s\hbar v_{\rm{F}} k$ is the massless dispersion and $k \doteq \abs{\bm k}$. The complete solution $\ket{\bm k s} \doteq \ket{\bm k} \otimes \ket{s}_{\bm k}$ is given by a plane wave $\ket{\bm k}$ multiplied by the spinors $\ket{s}_{\bm k}$, such that $\bra{\bm r}\ket{\bm k} = e^{i\bm r \cdot \bm k}/\sqrt{\mathcal A}$, with $\mathcal A$ the area of the graphene sample. The basis $\{\ket{\bm k s}\}$ forms a complete orthogonal set, thus verifying $\sum_{\bm k s} \ket{\bm k s}\bra{\bm k s} = \mathbb{1}$ and $\bra{\bm k s}\ket{\bm k' s'} = \delta_{\bm k \bm k'}\delta_{ss'}$. However, there is a nonzero overlap of pseudo-spinors, given by $v^{ss'}_{\bm k ,\bm k'} \doteq \leftidx{_{\bm k}}{\!\!\bra{s}}\ket{s'}_{\bm k'} = \big[1+ss'e^{i(\phi_{\bm k'}-\phi_{\bm k})}\big]/2$.

Let us now write the total Hamiltonian as
\begin{equation}
	\widehat{H}(t) = \widehat{H}_0 + \widehat{V}(t)\otimes\mathbb{1},
\end{equation}
where $\widehat{V}(t)$ is a diagonal potential on pseudo-spin space. We determine the potential by imposing its position representation, $V(\bm r,t) \doteq \bra{\bm r} \widehat{V}(t)\ket{\bm r}$, to verify the Poisson equation,
\begin{equation}
	\bm\nabla^2V = -\frac{e^2}{\varepsilon} n ,
\end{equation}
where $n$ is the total electronic density, $e$ is the elementary charge and $\varepsilon$ is the medium permittivity. A solution is found to be 
\begin{equation}
\widehat{V}(t) = \frac{e^2}{4\pi\varepsilon}\int d\bm r' \ \frac{n(\bm r',t)}{\abs{\widehat{\bm r}-\bm r'}}, \label{pot}
\end{equation}
where $\widehat{\bm r}$ is the position operator. This particular form for $\widehat{V}(t)$ is consistent with the initial single-particle approximation as it corresponds to the Hartree potential, thus discarding the contribution of the Fock and correlation terms. It is valid for small values of the coupling parameter $r_\text{s} \sim \alpha_\text{s} /\varepsilon_\text{r}$ (ratio of the average potential energy to the average kinetic energy), with $\alpha_\text{s} = e^2/(4\pi \varepsilon_0 \hbar v_\text{F}) \simeq 2.2$ the graphene structure constant, $\varepsilon_0$ the vacuum permittivity and $\varepsilon_\text{r}= \varepsilon/\varepsilon_0$. As a result, the Hartree approximation remains reliable as long as $\varepsilon_\text{r} \gtrsim 2.2$. 

To proceed, we deicide to work entirely with the density matrix $
\widehat \rho(t)$, instead of solving the Sch\"{o}dinger equation for the wave function. The density matrix evolves in time according to the quantum Liouville equation \cite{Liouville}, which reads
\begin{equation}
	\frac{\partial \widehat{\rho}}{\partial t} + \frac{i}{\hbar}[\widehat{H},\widehat{\rho}] = \mathcal S \{\widehat \rho \}, \label{liouville}
\end{equation}
with the right hand side (RHS) representing the collision operator. The density in Eq.~\eqref{pot} can then be written as the trace
\begin{equation}
	n(\bm r,t) = \tr[\delta(\bm r- \widehat{\bm r})\widehat{\rho}(t)], \label{density}
\end{equation}
and we recast Eq.~\eqref{liouville} in a more convenient form,
\begin{align}
	\frac{\partial}{\partial t}\rho_{\bm k \bm k'}^{ss'} =&- \frac{i}{\hbar}(\mathcal E_{\bm k}^s- \mathcal E_{\bm k'}^{s'})\rho_{\bm k \bm k'}^{ss'} - \frac{i}{\hbar}\sum_{\bm k'' s''}\Big( V_{\bm k \bm k''}^{ss''} \rho_{\bm k'' \bm k'}^{s''s'} \nonumber \\
	&- \rho_{\bm k \bm k''}^{ss''}V_{\bm k'' \bm k'}^{s'' s'}\Big) + \mathcal S_{\bm k \bm k'}^{ss'}, \label{matLiou}
\end{align}
where we have set $O_{\bm k \bm k'}^{ss'}\doteq \bra{\bm k s}{\widehat{O}(t)}\ket{\bm k' s'}$. 

\section{Wigner--Weyl formalism}
\label{sec_wigner}

A very handy way to treat the electronic system of Eq~\eqref{matLiou} is to use the Wigner--Weyl picture of quantum mechanics \cite{wigner,weyl}, which allows for a fully phase-space description, in close analogy with the classical case. In the classical limit, the Wigner function denotes the probability density of finding a particle in a given infinitesimal phase-space volume $d\bm rd\bm k$ centred in $(\bm{r},\bm{k})$. In the quantum case, due to the commutation relation between $\widehat{\bm{r}}$ and $\widehat{\bm{k}}$, the Heisenberg uncertainty principle prevents particles to localise in a specific phase-space point, and a proper distribution function (i.e., non-negative everywhere) is not possible \cite{moyal}. However, we can still construct the much renowned Wigner function $W(\bm{r}, \bm{k}, t)$, whose properties are similar to those of a classical distribution function, i.e., it is a function of both a spatial coordinate $\bm{r}$ and a wave-vector $\bm{k}$. Although $\bm{r}$ and $\bm{k}$ are classical-like variables and not conjugate quantum mechanical operators, they both give information about the spatial and momentum distributions of the system, which is described by a wave function $\psi(\bm{r},t)$. Consequently, we may have $W(\bm{r}, \bm{k}, t)<0$, with the phase-space regions where it takes negative values being purely quantum and having no classical analogue. For this reason, the Wigner function is often referred to as a quasiprobability function.

 The mathematical definition of the Wigner function is given by the Weyl transform of the density operator $\widehat{\rho}(t)$ \cite{weyl}, 
\begin{equation}
	W(\bm r,\bm k, t) = \int d\bm s \ e^{i\bm k \cdot \bm s} \bra{\bm r - \bm s/2} \widehat{\rho}(t)\ket{\bm r + \bm s/2}. \label{Wigner}
\end{equation}
The expectation value of any operator can be computed by integrating its Weyl transform multiplied by $W(\bm{r}, \bm{k}, t)$ over phase space, very much like in the classical case \cite{hillery}. Introducing the completeness relation for Dirac states $\ket{\bm k s}$, the Wigner function becomes
 \begin{equation}
	W(\bm r,\bm k, t) = \sum_{\bm q s s'} \ e^{i\bm q \cdot \bm r} \ v^{ss'}_{\bm k - \frac{\bm q}{2},\bm k +\frac{\bm q}{2}} \ \rho_{\bm k + \frac{\bm q}{2}, \bm k - \frac{\bm q}{2}}^{ss'}(t),
\end{equation}
from which we immediately get the Fourier transform 
\begin{equation}
	W(\bm q,\bm k,t) = \mathcal A\sum_{s s'} \ v^{ss'}_{\bm k - \frac{\bm q}{2},\bm k +\frac{\bm q}{2}} \ \rho_{\bm k + \frac{\bm q}{2}, \bm k - \frac{\bm q}{2}}^{ss'}(t).
\end{equation}
We stress the fact that $W(\bm q,\bm k,t)$ carries the dynamical information contained in the matrix elements of $\widehat\rho$, plus the lattice information contained in the overlap of the pseudo-spinors $v^{ss'}$. In the case of pure plane waves (e.g., a single-band massive plasma, $\mathcal E_{\bm k}\sim k^2$), the lattice information would be absent. By defining the matrix element $W^{ss'\!}(\bm q,\bm k,t) =\mathcal A v^{ss'}_{\bm k - \frac{\bm q}{2},\bm k + \frac{\bm q}{2}} \ \rho_{\bm k + \frac{\bm q}{2}, \bm k - \frac{\bm q}{2}}^{ss'}(t)$, we have
\begin{equation}
	W(\bm q,\bm k,t) = \sum_{ss'} W^{ss'\!}(\bm q,\bm k,t).
\end{equation}
When written in this manner, it becomes evident that the Wigner function correctly incorporates the dynamics of each band plus the contribution of the mixing terms, thus accounting for intra- and inter-band excitations. It is indeed known that the introduction of the spin (or pseudo-spin) makes us move to a tensorial formulation \cite{groot}. 

One can show that the total density, as defined in Eq.~\eqref{density}, has a simple form in terms of the Wigner function, namely
\begin{align}
	n(\bm r,t) &= \frac{1}{\mathcal A} \sum_{\bm k} W(\bm r,\bm k, t) = \sum_{ss'} n^{ss'\!}(\bm r,t), \label{densitW}
\end{align}
where $n^{ss'\!}(\bm r,t) \doteq \mathcal A^{-1}\sum_{\bm k} W^{ss'\!}(\bm r,\bm k,t)$. As discussed before, the diagonal elements correspond to the density of electrons in the conduction $(s=+1)$ and valence $(s=-1)$ bands, whereas the off-diagonal elements represent the coherence between both populations, i.e., the fraction of the electronic population that is constantly changing between the conduction and valence bands. 

Next, we derive an equation of motion for the Wigner elements in momentum space. Upon time differentiating $W^{ss'}$ and using Eq.~\eqref{matLiou}, we are lead to
\begin{widetext}
\begin{align}
	&i\hbar \frac{\partial}{\partial t}W^{ss'\!}(\bm q, \bm k,t) - \Delta \mathcal E^{ss'}_{\bm q \bm k} W^{ss'\!}(\bm q, \bm k,t) = v^{ss'}_{\bm k - \frac{\bm q}{2},\bm k +\frac{\bm q}{2}}\int \frac{d\bm q'}{(2\pi)^2} \ V(\bm q',t) \ \mathcal C\{W^{ss'}\} + i\hbar \mathcal A v^{ss'}_{\bm k -\frac{\bm q}{2},\bm k + \frac{\bm q}{2}}\mathcal S_{\bm k + \frac{\bm q}{2},\bm k - \frac{\bm q}{2}}^{ss'}, \label{wignerEq}
\end{align}
where $\Delta \mathcal E^{ss'}_{\bm q \bm k} \doteq \mathcal E_{\bm k + \frac{\bm q}{2}}^s -\mathcal E_{\bm k - \frac{\bm q}{2}}^{s'}$ and the kernel $\mathcal C\{W^{ss'}\}$ is defined as 
\begin{align}
\mathcal C\{W^{ss'}\} = \mathlarger{\sum}_{s''} \ \Bigg[& \frac{v^{ss''}_{\bm k + \frac{\bm q}{2},\bm k + \frac{\bm q}{2}-\bm q'}}{v^{s''s'}_{\bm k - \frac{\bm q}{2},\bm k + \frac{\bm q}{2}-\bm q'}} W^{s's''}(\bm q\! -\! \bm q',\bm k \!-\!\bm q'/2,t) - \frac{v^{s's''}_{\bm k - \frac{\bm q}{2}+\bm q',\bm k - \frac{\bm q}{2}}}{v^{s''s}_{\bm k - \frac{\bm q}{2}+\bm q',\bm k + \frac{\bm q}{2}}}W^{ss''}(\bm q \!-\! \bm q',\bm k \!+ \!\bm q'/2,t)\Bigg]
\end{align}
\end{widetext}
Moreover, we have used
\begin{equation}
	V_{\bm k \bm k'}^{ss'}(t) = \frac{1}{\mathcal A}v^{ss'}_{\bm k \bm k'}V(\bm k-\bm k',t),
\end{equation}
with $V(\bm k,t)$ the Fourier transform of $V(\bm r,t)$. By Fourier transforming Eq.~\eqref{pot}, we find $V(\bm k,t) = \mathcal U(\bm k) n(\bm k,t)$, with $\mathcal{U}(\bm k) = e^2/(2\varepsilon k)$ accounting for the (2D) time-independent Coulomb potential. 

We now consider small perturbations around an equilibrium configuration, keeping the lowest order contributions to the Wigner components. As such, we expand the density matrix as $\widehat{\rho}(t) = g_sg_v f(\widehat{H}_0 ) + \widehat{\delta \rho}(t)$, where $f(\mathcal E) = [e^{\beta(\mathcal E-\mu)}+1]^{-1}$ is the Fermi--Dirac distribution, $k_{\rm B}$ the Boltzmann constant, $T$ the absolute temperature, $\mu$ the chemical potential and $\beta=(k_{\rm B}T)^{-1}$. Besides, $g_sg_v$ accounts for the spin ($g_s=2$) and valley ($g_v=2$) degeneracy. The former results from the degeneracy of the spin populations in each energy band, which we have neglected in our treatment so far, and the latter should be incorporated to consistently include the two minima in the first Brillouin zone (BZ) \cite{pseudospin}. The perturbation $\widehat{\delta \rho}(t)$ is assumed to be small, $\widehat{\delta \rho}(t) \sim \order{r_\text{s}}$, such that higher powers in the coupling parameter are discarded. Consequently, the Wigner elements give
\begin{equation}
	W^{s s'}(\bm{q},\bm{k},t) \simeq g_sg_vf(\mathcal E_{\bm{k}}^s)\delta(\bm q)\delta_{ss'} + \delta W^{s s'}(\bm{q},\bm{k},t), \label{wignerPA}
\end{equation}
with $\delta W^{s s'}(\bm{q},\bm{k},t)$ a small perturbation. Similarly, the density and potential are perturbed as 
\begin{align}
n(\bm{q},t) &\simeq n_0 \delta(\bm{q}) + \delta n(\bm{q},t),\label{uu2} \\ 
V(\bm{q},t) &\simeq \mathcal{U}(\bm q)\left[ n_0 \delta(\bm{q}) + \delta n(\bm{q},t) \right] \label{uu2455} ,
\end{align}
where $n_0 =g_sg_v {\mathcal A}^{-1}\sum_{\bm k s}f(\mathcal E_{\bm k}^s)$ is the equilibrium density and $\delta n(\bm q,t)$ a small perturbation verifying $|\delta n(\bm q,t)| \ll 1$. Introducing Eqs.~\eqref{wignerPA}-\eqref{uu2455} into Eq.~\eqref{wignerEq}, and Fourier transforming the latter in the time domain, yields
\begin{align} 
	\delta W^{ss'}(\bm q,\bm k,\omega) = \ g_sg_v & \mathcal{U}(\bm q) \delta n(\bm q,\omega)\mathcal  F_{\bm k + \frac{\bm q}{2},\bm k - \frac{\bm q}{2}}^{ss'} \nonumber \\
	& \times \frac{f(\mathcal E_{\bm k - \frac{\bm q}{2}}^{s'}) - f(\mathcal E_{\bm k + \frac{\bm q}{2}}^{s})}{\hbar \omega-\mathcal E_{\bm k + \frac{\bm q}{2}}^s + \mathcal E_{\bm k - \frac{\bm q}{2}}^{s'}},\label{1storderW}
\end{align}
where $\mathcal F_{\bm k ,\bm k'}^{ss'} \doteq | v_{\bm k,\bm k'}^{ss'}|^2$ is the chirality factor, and a second-order term has been neglected, as well as the collision term. Equation \eqref{1storderW} is formally equivalent to Kubo's formula for the linear response of a many-body system \cite{walecka} and reproduces the features contained in the RPA for the collisionless limit $\mathcal S_{\bm k \bm k'}^{ss'}\rightarrow 0$ \cite{DiracPlasma, Wunsch_2006}. This formalism is specially advantageous to describe the dynamics of electrons that are far from equilibrium, such as the case of plasma instabilities, with the configuration being solely defined by the equilibrium distributions. Upon summing both sides of Eq.~\eqref{1storderW} over $\bm{k}$, $s$ and $s'$, we find the plasmon dispersion to be given by
\begin{equation}
\epsilon(\bm{q},\omega) = 0 ,
\label{Formaldisp}
\end{equation}
with $\epsilon(\bm{q},\omega) = 1 + \mathcal{U}(\bm q) \Pi(\bm{q},\omega)$ the dielectric function and $\Pi(\bm{q},\omega)$ the polarizability,
\begin{equation}
\Pi(\bm q, \omega) = \frac{g_sg_v}{\mathcal A}\mathlarger{\sum}_{\bm k s s'}\mathcal F^{ss'}_{\bm k,\bm k + \bm q} \ \frac{ f(\mathcal E_{\bm k +\bm q }^{s'})-f(\mathcal E_{\bm k}^s)}{\hbar\omega +\mathcal E_{\bm k }^s - \mathcal E_{\bm k + \bm q}^{s'} }.\label{POll}
\end{equation}
The formal result of Eq.~\eqref{Formaldisp} has been found before, in the context of the RPA \cite{Sarma2}. It applies in the weak-coupling limit $r_\text{s}\rightarrow 0$, where the plasmon frequency can be computed using the noninteracting irreducible polarizability. 

In what follows, we consider the case of negatively doped graphene, with the conduction band filled up to the Fermi level $\mathcal E_\text{F} = \hbar v_{\rm{F}} k_{\rm{F}} \simeq \mu \gg 0$. The Fermi level defines the Fermi wave-number $k_\text{F}$, and the latter is related with the doping density $n_0$ by 
\begin{equation}
k_\text{F} = \sqrt{\frac{4\pi n_0}{g_sg_v}},
\end{equation}
Typical experimental values of $n_0$ in the range $10^9 - 5\times 10^{12} ~\text{cm}^{-2}$ are achievable in graphene \cite{Adam}. In the case $\mathcal E_\text{F} \gg k_{\rm B}T$, the presence of holes (i.e., vacancies of valence electrons) is negligible, which allows us to set $f(\mathcal E_{\bm k +\bm q }^{-})-f(\mathcal E_{\bm k}^{-}) =0$ for all $\bm q$ and calculate the zeros of Eq.~\eqref{Formaldisp} explicitly (notice that the off-diagonal terms $s\neq s'$ in the polarizability are also not important since we are interested in the long-wavelength limit, for which we expand $\mathcal F_{\bm k, \bm k+\bm q}^{ss'} = \delta_{ss'} + \order{ q^2}$; simultaneously, the zeroth order in $\bm q$ of the remaining expression vanishes). For the sake of simplicity, we use the ultra-cold limit ($T\rightarrow 0$) of the Fermi--Dirac distribution,
\begin{equation}
f(\mathcal E^s_{\bm{k}}) \simeq \frac{n_0}{\pi k_\text{F}^2} \Theta(\mathcal E_\text{F} - \mathcal E_{\bm k}^s), \label{WignEQQ}
\end{equation}
with $\Theta(x)$ the Heaviside step function. After expanding Eq.~\eqref{Formaldisp} around $q = 0$ and keeping only terms up to $q^2$, the plasmon dispersion relation is obtained
\begin{equation}
\omega^2 = \omega_p^2 \frac{q}{k_\text{F}} + \frac{3}{4} v_\text{F}^2 q^2, \label{plasmon}
\end{equation}
where $\omega_p$ is the characteristic plasmon frequency,
\begin{equation}
\omega_p = \Bigg(\frac{e^2n_0v_\text{F}}{2\hbar \varepsilon}\Bigg)^{1/2}.
\end{equation}
For the typical experimental values $\varepsilon_\text{r} = 2.5$ and $n_0$ within $5\times10^{9}- 10^{12} \ \si{\centi\meter^{-2}}$, $\omega_p$ lies in the \si{\tera\hertz} region, between $2.6-37.3 \ \si{\tera\hertz}$. The first term $\omega\sim\sqrt{q}$ describes the long wavelength signature of plasmons in 2D electron gases \cite{kittel,finiteDoping,plasmon}. The most notable difference, when compared to the characteristic plasmon frequency in the 3D parabolic case, $\omega_p^{3D} = \sqrt{e^2n_0/(\varepsilon m)}$, is the appearance of $\hbar$ in leading order, revealing its pure quantum nature. Therefore, no classical counterpart exists for the 2D Dirac plasma. 
 
\begin{figure}[t!]
\hspace{-0.4cm}\includegraphics[scale=0.4]{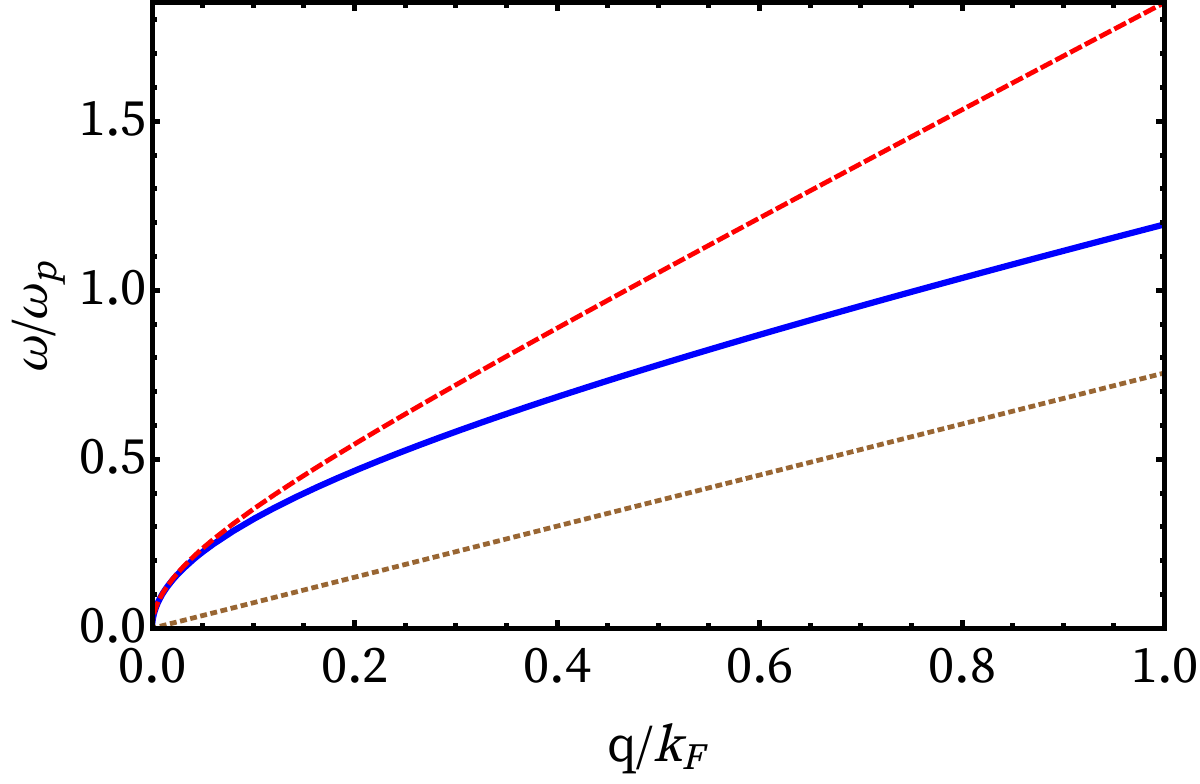}
\caption{Positive branch of the plasmon dispersion relation in ungated (solid blue) and gated (dashed red) configurations, along with the electron dispersion relation $\omega = v_\text{F} q$ (dotted brown) for $\varepsilon_\text{r}=2.5$ and $d_0k_\text{F} = 1$. Here, $q\doteq\abs{\bm q}$ is the wave-vector module.}
	\label{dispersionRelations}
\end{figure}

The same kinetic approach can be used to describe graphene electrons in a field effect transistor structure, i.e., placed between two metallic contacts, source and drain, and controlled by a gate. The gate voltage is related to the carrier density by \cite{capacitance} 
\begin{equation}
U(\mathbf{r},t) = e n(\mathbf{r},t) \left(\frac{1}{C_g} + \frac{1}{C_q} \right),
\label{voltage}
\end{equation}
where $C_{g}$ and $C_{q}$ are, respectively, the gate and quantum capacitance. The gate capacitance is given by $C_g = \varepsilon d_0$, where $d_0$ is the gate separation. The quantum capacitance $C_q$ reflects the change in the potential with the band occupancy and is defined as $C_q = e^2 D(\mathcal E)$, where $D(\mathcal E)=g_sg_v\abs{\mathcal E}/(2\pi \hbar^2v_\text{F}^2)$ is the density of states. For the typical carrier densities we are interested in, $C_g \ll C_q$, and so the second term in Eq.~\eqref{voltage} can be neglected. To include this effect in our model, one just needs to add $-eU(\mathbf{r},t)$ to the effective potential, which amounts to the substitution $\mathcal{U}(q) \rightarrow \mathcal{U}(q) + e^2 d_0/\varepsilon$. The dispersion relation becomes 
\begin{equation}
\omega^2 = \omega_p^2 \frac{q}{k_\text{F}} + \left(S^2+\frac{3}{4}v_\text{F}^2\right) q^2 +\mathcal{O}(q^3),
\label{gated}
\end{equation} 
with $S^2=2e^2v_\text{F}d_0\sqrt{n_0}/(\hbar \varepsilon \sqrt{\pi})$ and $S$ the sound velocity of the electron fluid \cite{graf4}. Equation \eqref{gated} is plotted in Fig. \ref{dispersionRelations}, alongside with Eq.~\eqref{plasmon} for comparison. It is patent that the gate induces a linear term $\omega\sim q $, which rapidly dominates over the $\sqrt{q}$ term, given a typical value for $k_\text{F} d_0$ of the order of unit. This is a consequence of the (static) screening produced by the gate potential.\\
To end this section, we adopt an approximation for Eq.~\eqref{wignerEq}, to simplify future calculations. As such, by realising that $V(\bm q,t)\sim 1/q$, it is convenient to evaluate the overlap factors on the RHS at $\bm q' \simeq 0$, since the integrand vanishes for very large $\bm q'$. Thus, Eq.~\eqref{wignerEq} will be approximated by
\begin{align}
	&i\hbar \frac{\partial}{\partial t}W^{ss'\!}(\bm q, \bm k,t) - \Delta \mathcal E^{ss'}_{\bm q \bm k}W^{ss'\!}(\bm q, \bm k,t) \nonumber \\
	&\simeq \int \frac{d\bm q'}{(2\pi)^2} \ V(\bm q',t) \Delta W^{ss'} + i\hbar \mathcal A v^{ss'}_{\bm k -\frac{\bm q}{2},\bm k + \frac{\bm q}{2}}\mathcal S_{\bm k + \frac{\bm q}{2},\bm k - \frac{\bm q}{2}}^{ss'}, \label{wignereqch4}
\end{align}
with $\Delta W^{ss'}(\bm q,\bm k,t;\bm q') \doteq W^{ss'}(\bm q- \bm q',\bm k -\bm q'/2,t)-W^{ss'}(\bm q - \bm q',\bm k+\bm q'/2,t)$. For the diagonal components $W^s \doteq W^{ss}(\bm r, \bm k ,t) $ the real-space (collisionless) version of Eq.~\eqref{wignereqch4} reads
\begin{align}
i\hbar \frac{\partial}{\partial t}W^{s} + i \hbar \mathcal K \{W^{s}\} = \int \frac{d\bm q}{(2\pi)^2} \ e^{i\bm{q}\cdot\bm{r}} \ \Big(&W^{s}_- - W^{s}_+\Big) \nonumber \\
&\times V(\bm{q},t), \label{realSpaceWig}
\end{align}
where $W^{s}_\pm \doteq W^s(\bm r, \bm k \pm \bm q/2,t)$ and
\begin{equation}
 \mathcal K\{W^{s}\} = sv_\text{F} \int d\bm r' \ \frac{ \sin \big(2\bm{k}\cdot \bm{r'}\big)}{\abs{\bm{r'}}^3} \ W^{s}(\bm{r}-\bm{r'},\bm{k},t) \label{kinj9}
\end{equation}
represents the kinetic operator. Its nonlocal nature is related to the relativistic character of Dirac electrons. In the classical limit $\hbar \rightarrow 0$, we can approximate $W_\pm^s$ as
\begin{equation} 
	W_{\pm}^s \simeq W^s \pm \frac{\hbar\bm q}{2}\cdot\frac{\partial W^s}{\partial \bm p} , \label{expans}
\end{equation}
which implies the RHS of Eq.~\eqref{realSpaceWig} to become
\begin{equation}
	\int \frac{d\bm q}{(2\pi)^2} \ e^{i\bm{q}\cdot\bm{r}} \ \Big(W^{s}_- - W^{s}_+\Big) \simeq i\hbar \frac{\partial V}{\partial \bm r} \cdot \frac{\partial W^s}{\partial \bm p},
\end{equation}
with $\bm p = \hbar \bm k$ the momentum. This does not seem to happen for the kinetic operator, which does not converge as $\hbar\rightarrow0$. However, by approximating $\sin (x)\simeq x$ in Eq.~\eqref{kinj9} (valid in the long-wavelength limit), one obtains $\mathcal K\{W^{s}\}\simeq \bm v^s_{\bm k}\cdot (\partial/\partial \bm r)W^s$, with $\bm v^s_{\bm k} \doteq \hbar^{-1}(\partial/\partial \bm k) \mathcal E_{\bm k}^s = sv_\text{F} \bm{k}/k$ the single-particle velocity. Under these approximations, we finally recover the (collisionless) Vlasov equation for the diagonal elements
\begin{equation}
	\frac{\partial W^s}{\partial t} + sv_{\rm{F}} \frac{\bm p}{p} \cdot \frac{\partial W^s}{\partial \bm r} - \frac{\partial V}{\partial \bm r} \cdot \frac{\partial W^s}{\partial \bm p}\simeq0, \label{vlsss}
\end{equation}
consistently with most semiclassical models present in the literature (see Ref.~\cite{stauber_2007} and references therein). It is convenient to make a comparison with the conventional parabolic dispersion relation, $\mathcal E^\text{par}_{\bm k} = \hbar^2 \bm k^2/(2m)$, for which the second term on the LHS of Eq.~\eqref{vlsss} would read \cite{hillery}
\begin{equation}
 \mathcal K\{W_\text{par}^{s}\}= \frac{\bm p}{m} \cdot \frac{\partial W^{s}_\text{par}}{\partial \bm r}, \label{Kparabolic}
\end{equation}
thus reducing to the usual (i.e., local) convective derivative. As a matter of fact, the Wigner--Weyl formulation of quantum mechanics could also be derived by applying quantum deformation theory to the classical Poisson algebra \cite{bizarro_2020}. 

\section{Fluid model}
\label{sec_hydrodynamics}

One of the major advantages of the present description is the possibility of computing the average values of operators, which are naturally given in terms of the Wigner function. Their time evolution can be computed using Eq.~\eqref{wignereqch4}, and a set of coupled equations can be put forward. As an example, we calculate the average momentum density 
\begin{equation}
\ov{\bm{\mathcal P}} (\bm r, t) = \tr\big[\widehat{\bm{\mathcal P}} \widehat \rho (t)\big],
\end{equation}
where
\begin{equation}
\widehat{\bm{\mathcal P}}=\frac{1}{2}\acomm{\widehat{\bm p}}{\delta(\bm r - \bm{\widehat r})},
\end{equation}
is the momentum density operator, $\widehat{\bm p} =\hbar \widehat{\bm k}$, and $\{,\}$ denotes symmetrisation. It follows that
\begin{equation}
\ov{\bm{\mathcal P}} (\bm r, t) = \frac{1}{\mathcal A} \sum_{\bm k ss'}\hbar \bm kW^{ss'}(\bm r,\bm k,t) \doteq \sum_{ss'}\ov{\bm{\mathcal P}}^{ss'\!\!}(\bm r, t),
\end{equation}
with the tensor $\ov{\bm{\mathcal P}}^{ss'\!\!}$ containing both the contribution of the particles in each band ($s=s'$) and of the particles which are in mixing states $(s\neq s'$). It is natural to define the average momentum $\ov{\bm p}^{ss'}$ through the relation $n^{ss'}\ov{\bm p}^{ss'\!\!} =\ov{\bm{\mathcal P}}^{ss'\!\!}$, thus setting the relevant macroscopic variables to be
\begin{equation}
	n^{ss'}(\bm r, t) = \int \frac{d\bm k}{(2\pi)^2} \ W^{ss'\!}(\bm r, \bm k ,t) \label{nn} 
\end{equation}
and
\begin{equation}
	\ov{\bm p}^{ss'} (\bm r, t) = \frac{1}{n^{ss'}}\int \frac{d\bm k}{(2\pi)^2} \ \hbar \bm k \ W^{ss'\!}(\bm r, \bm k ,t) ,\label{pp} 
\end{equation}
after replacing the sums with integrals. \\

Henceforth, and in order to simplify our analysis, we neglect the off-diagonal contributions, in as much as we are interested in the semiclassical limit of negatively doped graphene, for which case they play a minimal role. We are, thus, allowed to drop the double index to facilitate the notation (as, e.g.,  $W^{s}\doteq W^{ss}$). The time evolution equations for the diagonal quantities in Eqs.~\eqref{nn} and \eqref{pp} provide (see Appendix \ref{apA})
\begin{equation}
\frac{\partial n^{s}}{\partial t} + \frac{\partial}{\partial \bm r}\cdot \ov{\bm{j}}^s =0 \label{nnSep1}
\end{equation}
and 
\begin{equation}
\frac{\partial }{\partial t}\left( n^s\ov{\bm{p}}^{s}\right) + \frac{\partial P^s}{\partial \bm r} + n^{s}\frac{\partial V}{\partial \bm r} = 0, \label{kkSep1}
\end{equation}
where $\ov{\bm j}^s=n^s \ov{\bm v}^s + \ov{\bm j}_{\rm Q}^s$ is the density current, comprising the usual classical term  
\begin{equation}
	n^s\ov{\bm{v}}^s = \int \frac{d\bm k}{(2\pi)^2} \ \bm v_{\bm k}^s \ W^{s}(\bm{r},\bm{k},t), \label{avVel}
\end{equation}
plus a quantum ($\hbar$-dependent) part
\begin{align}
	\ov{\bm j}_{\rm Q}^s = \sum_{n \in \! \ \text{odd}}& \left(\frac{\hbar}{i}\right)^{n+1} \int\frac{d\bm k d\bm q}{(2\pi)^4}\frac{ W^{s}(\bm{q},\bm{k},t)}{(n+2)!}\nonumber \\
	&\times \Bigg[e^{i\bm q\cdot \bm r}\left(\frac{1}{2}\overleftarrow{\frac{\partial}{\partial \bm r}}\cdot \overrightarrow{\frac{\partial}{\partial \bm p}}\right)^{n+1} \bm v_{\bm p}^s\Bigg], \label{jQuantum}
\end{align}
stemming from the non-vanishing $\bm p$-derivatives of the Dirac velocity, $\bm v_{\bm p}^s \doteq sv_{\rm F}\bm p /p$. The pressure tensor takes the form $P^s = P^s_{\rm cl} + P_{\rm Q}^s$, where 
\begin{equation}
[P^{s}_{\rm cl}]_{lm}= sv_\text{F} \int \frac{d\bm k}{(2\pi)^2} \ \frac{p_lp_m}{p} \ W^s(\bm{r}, \bm{k}, t) , \label{prrrs}
\end{equation}
is the 2D (classical) pressure, and $P_{\rm Q}^s$ introduces quantum-mechanical contributions,
\begin{align}
	[P_{\rm Q}^s]_{lm} =& \sum_{n \in \! \ \text{odd}} \left(\frac{\hbar}{i}\right)^{n+1} \int\frac{d\bm k d\bm q}{(2\pi)^4} \frac{ W^{s}(\bm{q},\bm{k},t)}{(n+2)!} p_l \nonumber \\
	&\times \Bigg[e^{i\bm q\cdot \bm r}\left(\frac{1}{2}\overleftarrow{\frac{\partial}{\partial \bm r}}\cdot \overrightarrow{\frac{\partial}{\partial \bm p}}\right)^{n+1} [v_{\bm p}^s]_m\Bigg]. \label{quantumCorr}
\end{align}
Above, $[v_{\bm p}^s]_m$ refers to the $m$-th component of $\bm v_{\bm p}^s$. By taking the limit $\hbar\rightarrow 0$, it follows that $\ov{\bm j}^s\rightarrow n^s \ov{\bm v}^s$ and $P^s \rightarrow P^s_{\rm cl}$, as expected for the classical case. The classical limit of Eqs.~\eqref{nnSep1} and \eqref{kkSep1} would thus be recovered if one had replaced the Wigner equation by the Vlasov equation from the start. Note that the new terms arise after taking into account not only the local contribution ($\sim \bm v_{\bm k}^s$), but all (non-local) terms of Eq.~\eqref{kinj9}. To the best of our knowledge, Eqs.~\eqref{jQuantum} and \eqref{quantumCorr} have not been given anywhere else. In the case of systems with parabolic dispersion, $\ov{\bm j}_{\rm Q}^s$ and $P_{\rm Q}^s$ would vanish and the fluid equations derived from the Wigner model would present no advantage over its classical analogue. This can be deduced from Eqs.~\eqref{jQuantum} and \eqref{quantumCorr} by simply assuming the dependence $\bm v_{\bm k}^s \sim \bm k$. A reason for this particular behavior relies on the purely quantum structure of Eq.~\eqref{CVhamiltonian}, in contrast to the kinetic term for massive particles, which takes the same form irrespective of treating the classical or the quantum case (with the momentum being, respectively, a classical variable or a quantum operator). On the contrary, the Dirac kinetic operator has no classical analog.

By including the first terms of $\ov{\bm j}_{\rm Q}^s$ and $P_{\rm Q}^s$, we obtain the modified fluid equations 
\begin{align}
\frac{\partial n^{s}}{\partial t} + \frac{\partial}{\partial \bm r}\cdot &(n^s\ov{\bm v}^s) = \frac{\hbar^2}{24} \Bigg[ \frac{\partial^3}{\partial x^3} \Big(n^s\ov{J_{xxx}}^s \Big) \nonumber \\
&+ 3\frac{\partial^2}{\partial x^2}\frac{\partial}{\partial y} \Big(n^s\ov{J_{xxy}}^s \Big) +(x\leftrightarrow y)\Bigg] ,\label{finalnn}
\end{align}
and 
\begin{align}
\frac{\partial }{\partial t} (n^s\ov{\bm{p}}^{s}) & + \frac{\partial P^s_{\rm cl}}{\partial \bm r} + n^{s}\frac{\partial V}{\partial \bm r} = \frac{\hbar^2}{24} \Bigg[ \frac{\partial^3}{\partial x^3} \Big(n^s\ov{\bm{T}_{xxx}}^s \Big) \nonumber \\
&+ 3\frac{\partial^2}{\partial x^2}\frac{\partial}{\partial y} \Big(n^s\ov{\bm{T}_{xxy}}^s \Big) + (x \leftrightarrow y) \Bigg]. \label{finalkk}
\end{align}
The tensors $J^s_{ijl}$ and $\bm T^s_{ijl}$ result from the first ($\sim \hbar^2$) terms of $\ov{\bm j}_{\rm Q}^s$ and $P_{\rm Q}^s$, and read 
\begin{equation}
	J^s_{ijl} = sv_\text{F}\Bigg(\frac{3p_ip_jp_l}{\abs{\bm{p}}^5} -\frac{\delta_{ij} p_l + \delta_{jl}p_i + \delta_{li} p_j }{\abs{\bm{p}}^3} \Bigg), \label{jhbar} 
\end{equation}
and 
\begin{equation}
	\bm T^s_{ijl} = \bm p J^s_{ijl} \label{Thbar}.
\end{equation}
Furthermore, $\ov{G}^s(\bm r,t)$ denotes the average value of some generic function $G^s(\bm p)$, akin to Eq.~\eqref{avVel}. We will provide explicit expressions for these quantities below and discuss how they modify the plasmon dispersion relation.

%%%%%%%%%%%%%%%%%%%%%%%%%%%%%%%%%%%%%%%%%%%%%%%%%%%%%%%%%%%%%%%%%%%%%
\subsection{Calculation of the macroscopic variables}

Having set the relevant transport equations, we move now to a more detailed discussion concerning the averaged momentum and velocity fields. Being interested in the case of full valence band and doped conduction band ($\mu \gg k_{\rm B} T$), we concentrate on the conduction electrons only, thus dropping the band index ($s=1$). As one can readily observe by comparing Eqs.~\eqref{pp} and \eqref{avVel}, a proportionality relation of the form $\ov{\bm p} = m\ov{\bm v}$ does not hold for massless particles, with a constant $m$ \cite{graf6}. However, by allowing a space and time dependence on the mass, we are able to define an effective ``mass tensor'', $m_{ij}\doteq m_{ij}(\bm r ,t)$, as
\begin{equation}
 \ov{p}_i =  \sum_j  m_{ij}\ov{v}_j , \label{mass}
\end{equation}
The meaning of such field should be clear: although the carriers have no mass, since they are Dirac particles, the fluid velocity and momentum fields can be related via an effective mass, providing a measure of the inertia of a fluid element. A tensorial structure such as that of Eq.~\eqref{mass} is required since, at a first glance, we should anticipate the components of $\ov{\bm p}$ in a given direction $i$ to be given by a linear combination of the velocity components in each direction $j$, with coefficients $m_{ij}$. Nevertheless, if one is dealing with a rotationally invariant system, the mass tensor becomes diagonal, $m_{ij}= \delta_{ij}m$. \\

Let us now modify the equilibrium used in the last section in order to accommodate for a (possibly local) average velocity of the electronic fluid, namely,
\begin{equation}
	W(\bm{r},\bm{k},t) =  \frac{g_sg_v}{1+e^{\beta(\mathcal E_{\bm k} - \hbar \bm k \cdot \bm u - \mu)}},\label{Deformed}
\end{equation}
where $\mu \doteq \mu(\bm r,t)$, $\bm u\doteq \bm u(\bm r,t)$ and $\beta \doteq 1/[k_{\rm B}T(\bm r,t)]$ are local variables to be calculated by imposing the conservation of number of particles, momentum and energy, respectively \cite{conserved}. Additionally, this particular choice for the Wigner function cancels the collision term on the Wigner equation \cite{graf7,rapidc}. Computations with  Eq. \eqref{Deformed} are valid under the assumptions that the macroscopic quantities are slow-varying in both space and time, which is the natural requirement to go from a microscopic field theory to a macroscopic fluid model. The fluid description is valid if changes in the macroscopic quantities take place on large spacial and temporal scales, i.e., if $q\ll k_\text{F}$ and $\omega\ll \omega_p$, respectively. This requirement is fulfilled if the characteristic time of the kinematic processes, $t\sim 1/\omega$, is much longer than the inverse collision frequency, $1/\nu_c$, and the typical length $L$ is much greater than the mean free path $l\sim v_\text{F}/\nu_c$, so that the plasma can be regarded locally as in a quasi-equilibrium configuration. However, due to the quantum nature of the model, we were able to capture the relativistic Dirac structure in a rigorous way. Actually, in the context of 2D quantum plasmas achieved in semiconductor structures, the De Broglie wavelength is replaced by the Thomas-Fermi screening length $\lambda_{\rm TF} = 2\pi/(g_sg_v r_\text{s}k_\text{F})$. This is the analogue of the classical Debye length in plasmas, and differs from the Fermi wavelength, $\lambda_\text{F} = 2\pi/k_\text{F}$ ($\sim 10~{\rm nm}-10 \mu{\rm m}$ for typical graphene parameters). Consequently, the fluid limit is expected to be valid provided the conditions
\begin{equation}
q\lambda_{\rm TF}\ll 1, \quad q\lambda_\text{F}\ll 1
\end{equation}
apply. In graphene, $\lambda_\text{F}/\lambda_{\rm TF} \simeq 8.8/\varepsilon_\text{r}$, such that the second condition is the most stringent. Typical values of $k_\text{F}$ are found between $10^3$ and $10^6$ $\text{cm}^{-1}$. At smaller wavelengths, the microscopic structure becomes important and the fluid approximation no longer holds. \\
As expected, Eq.~\eqref{avVel} provides $\ov{\bm v} = \bm u$. In its turn, Eqs.~\eqref{nn} and \eqref{pp} yield
\begin{equation}
	n(\bm r,t) = \frac{n_0}{(1-\xi^2)^{\frac{3}{2}}}
\end{equation}
and
\begin{equation}
	\ov{\bm p}(\bm r,t) = m(\bm r,t) \ov{\bm v},
\end{equation}
where $\xi(\bm r,t) \doteq \ov{v}(\bm r,t)/v_{\rm{F}}$ is the reduced velocity, $n_0$ is the density at rest,
\begin{equation}
	n_0 = -g_sg_v\text{Li}_{2}(-e^{\beta \mu})\frac{1}{2\pi \beta^2\hbar^2v_{\rm F}^2},
\end{equation}
and 	
\begin{equation}
	m(\bm r,t) = -g_sg_v\text{Li}_{3}(-e^{\beta \mu})\frac{3}{2\pi \beta^3\hbar^2 v_{\rm F}^4n_0}\frac{1}{1-\xi^2}
\end{equation}
is the local mass. $\rm{Li}_n$ denotes the polylogarithm function of order $\rm n$. Both $m$ and $n_0$ admit amenable forms in the limit $T\rightarrow 0$, in which case we find $n_0  \simeq \pi k_{\rm F}^2$ and
\begin{equation}
m  \simeq \frac{\mathcal M}{1-\xi^2}, \label{masssmall}
\end{equation}
with $\mathcal M = \hbar k_{\rm F}/v_{\rm F}$ being Drude's mass. Note that this result differs from that which can be found in the literature \cite{graf7,cgraf7}. The classical pressure reduces to 
\begin{equation}
	P_{\rm cl}(\bm r,t) = P_0(f_1\mathbb{1} + f_x\sigma_x + f_z\sigma_z),
\end{equation}
with
\begin{equation}
f_1= \frac{1+\frac{\xi^2}{2}}{(1-\xi^2)^{\frac{5}{2}}},
\end{equation}
\begin{equation}
f_x = \frac{3}{2}\frac{\xi^2\cos2\Omega}{(1-\xi^2)^{\frac{5}{2}}},
\end{equation}
and
\begin{equation}
f_z = \frac{3}{2}\frac{\xi^2\cos^2\Omega}{(1-\xi^2)^{\frac{5}{2}}}.
\end{equation}
In the above expressions, $\Omega = \arctan(\ov{v}_y/\ov{v}_x)$ is the polar angle of the fluid velocity and 
\begin{equation}
	P_0 = -g_sg_v\text{Li}_{3}(-e^{\beta \mu})\frac{1}{2\pi^2 \beta^3\hbar^2v_{\rm F}^2}
\end{equation}
is the pressure in the absence of flow. The $T\rightarrow 0$ limit renders $P_0 \simeq p_{\rm F}v_{\rm F}n_0/(3\pi^2)$. The average value of the quantum tensor $J_{ijl}$ vanishes under the assumption of Eq.~\eqref{Deformed}, i.e., $\ov{J_{ijl}}=0$ for any $\{ijl\}$, which yields the density current to be approximated by its classical value up to second order in $\hbar$, $\ov{\bm j}  \simeq n \ov{\bm v}$. On the contrary, we find a non-zero contribution from the average value of $\bm T_{ijl}=(T_{ijl}^x,T_{ijl}^y)=(p_x J_{ijl},p_yJ_{ijl})$. Given the symmetry properties of Eqs.~\eqref{jhbar} and \eqref{Thbar}, the independent components are found to be
\begin{align}
	\ov{T_{xxx}^y}&= \frac{3T_0}{2}(g_2 \cos 2\Omega + g_3)\sin2\Omega,\\
	\ov{T_{xxy}^x}&= \frac{T_0}{2}(3g_2 \cos 2\Omega - g_3)\sin2\Omega,\\
	\ov{T_{xxy}^y}&= \frac{T_0}{4}(-g_1 -3g_2 \cos 4\Omega - 4g_3\cos 2\Omega),\\
	\ov{T_{yyx}^x}&= \frac{T_0}{2}(-g_1 -3g_2 \cos 4\Omega + 4g_3\cos 2\Omega),\\
	\ov{T_{yyx}^y}&= \frac{T_0}{2}(-3g_2 \cos 2\Omega - g_3)\sin2\Omega,\\
	\ov{T_{yyy}^x}&= \frac{3T_0}{2}(-g_2 \cos 2\Omega + g_3)\sin2\Omega,\\
	\ov{T_{xxx}^x}&= \ov{T_{yyy}^y}= \frac{3T_0}{4}(-g_1 + g_2 \cos 4\Omega),
\end{align}
with the auxiliary functions defined as
\begin{align}
	g_1 =& \frac{1}{(1-\xi^2)^{\frac{1}{2}}},\\
	g_2=& \frac{8(1-\xi^2)+\xi^4+4(\xi^2-2)(1-\xi^2)^{\frac{1}{2}}}{\xi^4(1-\xi^2)^{\frac{1}{2}}},\\
	g_3 =& \frac{\xi^2-2+2(1-\xi^2)^{\frac{1}{2}}}{\xi^2(1-\xi^2)^{\frac{1}{2}}},
\end{align}
and 
\begin{equation}
	T_0 = -g_sg_v\text{Li}_{1}(-e^{\beta\mu})\frac{1}{4\pi n\beta\hbar^2}.
\end{equation}
%%%%%%%%%%%%%%%%%%%%%%%%%%%%%%%%%%%%%%%%%%%%%%%%%%%%%%%%%%%%%%%%%%%%%
\subsection{Semiclassical dispersion relation}

\begin{figure}[t!]
\hspace*{-0.5cm}
\includegraphics[scale=0.4]{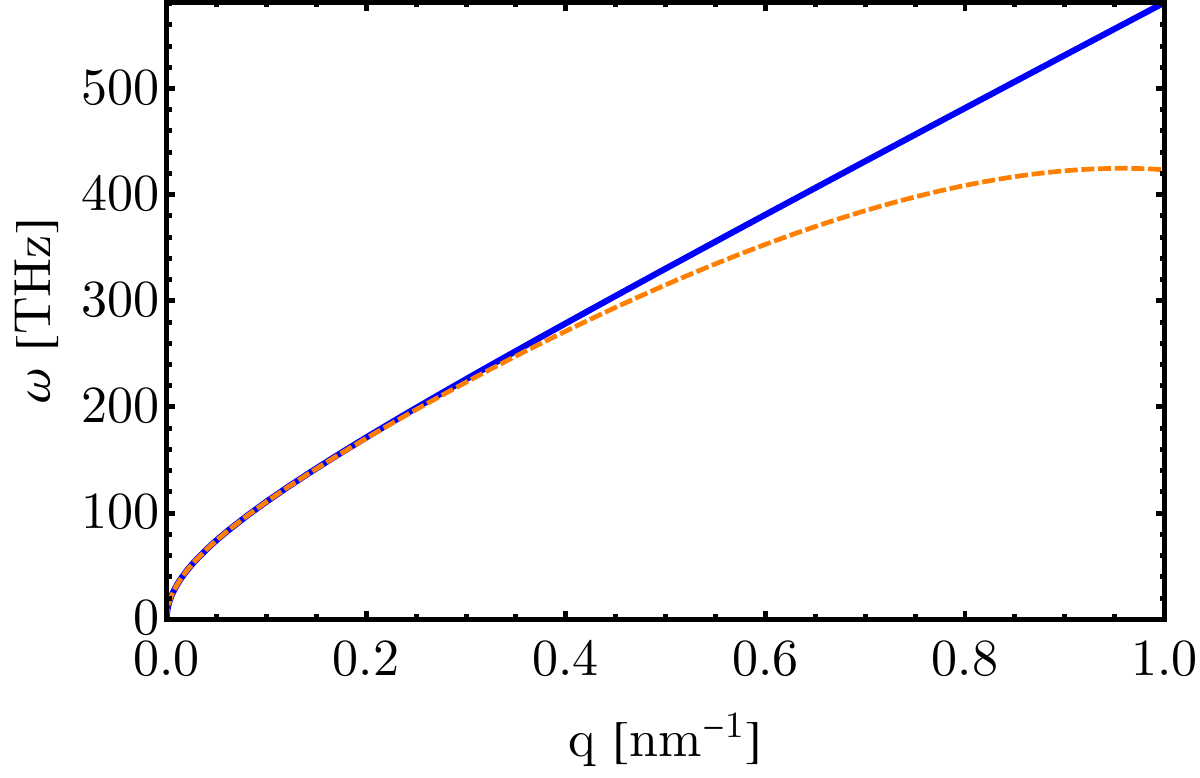}
\caption{Positive branch of the semiclassical plasmon dispersion relation in Eq.~\eqref{quantPlasmonn} (orange dashed), together with its classical counterpart, given by Eq.~\eqref{plasmon} (solid blue), with $\varepsilon_\text{r}=2.5$.}
\label{pla}
\end{figure}

Next, we linearize Eqs.~\eqref{finalnn} and \eqref{finalkk} to find the modified dispersion relation, which is valid up to $\hbar^2$, provided the higher order terms can be neglected. The case of interest is, again, the conduction band and we restrict the variations to the $x-$direction, which implies that $\Omega=0$ and the relevant quantities to be $[P_{\rm cl}]_{xx}$, $\ov{J_{xxx}}$ and $\ov{T^x_{xxx}}$. For small fluid velocity, $\xi \ll 1$, all fluid variables excluding $\ov{p}_x$ can be expanded to first order in $\xi^2$, which brings Eqs.~\eqref{finalnn} and \eqref{finalkk} to its linearised form (see Appendix \ref{apB})
\begin{equation}
	\frac{3}{2}\frac{\partial \xi^2}{\partial t} + v_{\rm F} \frac{\partial \xi}{\partial x} = 0, \label{qc1}
\end{equation}
and 
\begin{equation}
\mathcal M v_{\rm F}\frac{\partial \xi}{\partial t} + \frac{3}{2}\frac{p_{\rm F}v_{\rm F}}{\pi ^2} \frac{\partial \xi^2}{\partial x} + \frac{\partial \delta V}{\partial x} = -\frac{1}{16\pi^2}\frac{\hbar ^2}{4 \mathcal M} \frac{\partial^3 \xi^2}{\partial x^3} , \label{qc2}
\end{equation}
where
\begin{equation}
	\delta V(\bm r,t) = \frac{e^2n_0}{4\pi \varepsilon } \int d\bm r' \ \frac{1+\frac{3}{2}\xi^2}{\abs{\bm{r}-\bm{r'}}}, \label{qc3}
\end{equation}
is the perturbed potential. After Fourier transforming the system of differential equations~\eqref{qc1}$-$\eqref{qc2}, it reduces to an algebraic system for two unknowns, $\xi(\bm q,t)$ and $\xi^2(\bm q,t)$. Note that $\xi^2(\bm q,\omega)$ is not the square of $\xi(\bm q,\omega)$, but rather the Fourier transform of the square of $\xi(\bm r,t)$. From the two equations we can retrieve the dispersion relation of the plasma waves,
\begin{equation}
	\omega^2= \omega_p^2 \frac{q}{k_\text{F}} + \frac{3}{4} v_\text{F}^2 q^2 -\frac{1}{16}\left(\frac{\hbar^2 q^4}{4 \mathcal{M}^2}\right).
	\label{quantPlasmonn}
\end{equation}
The first two terms on the RHS are in agreement with Eq.~\eqref{plasmon}. The third one accounts for a quantum correction ($\sim \hbar^2$) known as the Bohm term \cite{bohmfinal}, which is modified by a small numerical factor. The Bohm potential plays the role of a quantum pressure and here results from the contribution of $\ov{T^x_{xxx}}$. For Dirac plasmas, we find a negative sign in the quantum correction (the Bohm pressure softens the plasmon mode), contrary to the what has been reported for the case of parabolic fermions, in which case it takes the form $\hbar^2q^4/(4m^2)$, with $m$ being the usual (constant) mass \cite{DensePlasmas1,hydro1}. Since the magnitude of this correction is small, it only becomes important for intermediate wave-number values, as can be seen in Fig. \ref{pla}. The plasmon dispersion relation of Eq.~\eqref{quantPlasmonn} indicates that the present quantum fluid model goes beyond the RPA calculation performed in previous sections, as the latter hides the effect of the Bohm dispersion, which is known to play an important role in the case of dense plasmas \cite{DensePlasmas2}. \par 
%%%%%%%%%%%%%%%%%%%%%%%%%%%%%%%%%%%%%%%
\section{Conclusions and outlook} 
\label{sec_conclusions}

Using the Wigner--Weyl description, we have derived a quantum fluid model for a Dirac plasma in single-layer graphene. The massless nature of the quasi-particles was captured by the low-energy limit of the many-body Hamiltonian, which served as the basis for the construction of the Wigner matrix. Then, the equation of motion for the Wigner matrix elements in phase space was established, with the interacting potential being given by the Hartree approximation. The latter turns out to be an excellent approximation for graphene, as the coupling constant is very small, $r_s \sim \alpha_s/\varepsilon_\text{r} \ll 1$, and independent of the electron density. Additionally, a closed set of generalized quantum fluid equations was derived by taking the moments of the quantum kinetic equation. We found an infinite number of $\hbar$-dependent terms arising in both the continuity and force equations as a consequence of the linear dispersion relation of the carriers, and that introduced corrections to the current density and pressure. As it is shown, in the conventional case of parabolic systems (i.e. for which the single-particle dispersion relation is $\sim p^2$), these contributions vanish. To the best of our knowledge, such corrections have not been discussed so far in the literature. By neglecting the quantum corrections, our model reduces to existing semiclassical models for Dirac fermions \cite{graf3,graf1}. Moreover, a nonlinear relation between the fluid velocity and momentum fields has been put forward in order to construct a hydrodynamical mass. For small fluid velocity, we found a relation of the form $\ov{\bm p} = \gamma \mathcal{M}\ov{\bm v}$, where $\mathcal M$ is Drude's mass and $\gamma \simeq (1+\ov{v}^2/v^2_{\rm F})$ is a local Lorentz-like factor. We stress the fact that a quantitative description for the hydrodynamical mass is a major issue in the current research, therefore making our findings on the hydrodynamical mass an important result of the current kinetic description. We have also computed higher-order corrections to the plasmon dispersion relation ($\sim q^4$) by taking into account the first quantum corrections to the fluid equations.\par 

We expect our formalism to be particularly suited to describe out-of-equilibrium quantum dynamics of Dirac electrons in a more systematic manner. While the linear response can be indeed described within the framework of the RPA (which we recover upon linearization of the equations for the Wigner function), the nonlinear regimes deserve a more complete description, which are captured by the kinetic equation that we have proposed here. This is crucial in the description of saturation in the late stages of dynamical instabilities, for example. Moreover, (quantum) diffusive processes may be captured by adapting the strategy of quasilinear theories accounting for particle-wave interactions \cite{nicholson_1983}; by working out the collision integral at different levels of approximation, we may obtain quantum versions of the Fokker-Planck equation for relativistic-like particles. Another interesting aspect of our work is the contribution to the hydrodynamical modeling of graphene plasmas. With the \textit{ab initio} establishment of a hydrodynamical mass, we are able to simulate experimentally relevant configurations, such as hydrodynamical instabilities in field-effect transistor THz emitters \cite{graf4}. Moreover, at a phenomenological level, our hydrodynamical model is particularly suited to accommodate both shear and odd (or Hall) viscosity terms. In the presence of strong magnetic fields, the Chapman-Enskog approach to hydrodynamics yields a term in the fluid equations of the form $\sim\nu_o \nabla^2 \epsilon_{ij} v_j$, where $\nu_o$ is some function of the magnetic field and temperature and $\epsilon_{ij}$ is the Levi-Civita symbol \cite{avron_1998}. The appearance of $ \epsilon_{ij}$ term makes the viscosity tensor off-diagonal and, thus, dissipation free, contrary to what happens in the case of even viscosity. Odd viscosity has been identified to play a special role in the topology of acoustic waves \cite{souslov}, and we hope that plasmons may also feature such topological transitions, opening the venue for a range of applications. \\

\acknowledgments

H. T. acknowledges Funda\c{c}\~{a}o para a Ci\^{e}ncia e a Tecnologia (FCT-Portugal) through the Contract No. CEECIND/00401/2018 and through the exploratory project No. UTA-EXPL/NPN/0038/2019. J. P. S. B. acknowledges the financial support from the Quantum Flagship Grant PhoQuS under Grant No. 820392 of the European Union. The Authors further thank the anonymous reviewers, whose comments and criticism very much helped to improve our manuscript.

%%%%%%%%%%%%%%%%%%%%%%%%%%%%%%%%%%%%%%%%%%%%%%%%%%%%%%%%%%%%%%%%%%%%%

\onecolumngrid
\begin{appendices}
\section{Derivation of the fluid set of equations}\label{apA}
In this appendix, we derive the set of Eqs.~\eqref{nn1} and \eqref{pp1}. To do so, we start by time differentiating Eqs.~\eqref{nn} and \eqref{pp} (multiplied by $n^s$), and replace $(\partial/\partial t)W^{s}(\bm{r},\bm{k},t)$ by $(2\pi)^{-2}\int d\bm q \ e^{i\bm{q}\cdot\bm{r}} \ (\partial/\partial t) W^{s}(\bm{q},\bm{k},t)$, given that Eq.~\eqref{wignereqch4} only provides time derivatives of the Wigner matrix components in Fourier space. We get
\begin{align}
 \frac{\partial n^{s} }{\partial t}= & -\frac{1}{\hbar}\int\frac{d\bm k d\bm q}{(2\pi)^4} \ ie^{i\bm{r}\cdot\bm{q}} \ \Delta \mathcal E^{ss}_{\bm q \bm k}\ W^{s}( \bm{q}, \bm{k}, t) -\frac{1}{\hbar} \int \frac{d\bm qd\bm q'}{(2\pi)^4} \ ie^{i\bm{q}\cdot\bm{r}} V(\bm{q'},t) \int \frac{d\bm k}{(2\pi)^2}\ \Delta W^{ss} \label{dtnn}
\end{align}
and
\begin{align}
\frac{\partial }{\partial t} \big(n \ov{\bm{p}}^{s}\big) =& - \int \frac{d\bm k d\bm q}{(2\pi)^4} \ ie^{i\bm{r}\cdot\bm{q}} \ \bm{k} \ \Delta \mathcal E^{ss}_{\bm q \bm k} \ W^{s}( \bm{q}, \bm{k}, t) - \int\frac{d\bm qd\bm q'}{(2\pi)^4} \ i e^{i\bm{q}\cdot\bm{r}} \ V(\bm{q'},t) \int \frac{d\bm k}{(2\pi)^2} \ \bm{k} \ \Delta W^{ss}. \label{dtkk}
\end{align}
Note that the contribution of collisions vanishes after requiring the conservation of the number of particles and the total momentum before and after each collision. Next, we plug the identities
\begin{align}
 \Delta W^{ss} = &-2 e^{-\bm{q'}\cdot \frac{\partial}{\partial \bm q}}\sinh(\frac{\bm{q'}}{2}\cdot \frac{\partial}{\partial \bm k}) W^{s} (\bm{q},\bm{k},t)
 \end{align}
and 
 \begin{align}
 \Delta \mathcal E^{ss}_{\bm q \bm k} = & \ 2\sinh(\frac{\bm{q}}{2}\cdot\frac{\partial}{\partial \bm k}) \mathcal E^s_{\bm k}
\end{align}
into Eqs.~\eqref{dtnn} and \eqref{dtkk}, which leads to 
\begin{align}
 \frac{\partial n^{s}}{\partial t} =& -\frac{2}{\hbar}\int \frac{d\bm kd\bm q}{(2\pi)^4}\ ie^{i\bm{r}\cdot\bm{q}} \ \Bigg[\sinh(\frac{\bm{q}}{2}\cdot \frac{\partial}{\partial \bm k})\mathcal E^s_{\bm k}\Bigg] W^{s}( \bm{q}, \bm{k}, t) \nonumber \\
 & +\frac{2}{\hbar} \int \frac{d\bm q d\bm q'}{(2\pi)^4} \ ie^{i\bm{q}\cdot\bm{r}} V(\bm{q'},t)e^{-\bm{q'}\cdot \frac{\partial}{\partial \bm q}} \int \frac{d\bm k}{(2\pi)^2}\ \sinh(\frac{\bm{q'}}{2}\cdot \frac{\partial}{\partial \bm k}) W^{s} (\bm{q},\bm{k},t) \label{n2} 
 \end{align}
 and
\begin{align}
\frac{\partial }{\partial t} \big(n^s\ov{\bm{p}}^{s}\big)= & -2\int \frac{d\bm k d\bm q}{(2\pi)^4}\ ie^{i\bm{r}\cdot\bm{q}} \bm k\ \Bigg[\sinh(\frac{\bm{q}}{2}\cdot \frac{\partial}{\partial \bm k})\mathcal E^s_{\bm k}\Bigg] W^{s}( \bm{q}, \bm{k}, t) \nonumber \\
	 & +2 \int \frac{d\bm q d\bm q'}{(2\pi)^4} \ ie^{i\bm{q}\cdot\bm{r}} V(\bm{q'},t)e^{-\bm{q'}\cdot \frac{\partial}{\partial \bm q}} \int \frac{d\bm k}{(2\pi)^2}\ \bm k \sinh(\frac{\bm{q'}}{2}\cdot \frac{\partial}{\partial \bm k}) W^{s} (\bm{q},\bm{k},t). \label{p2}
\end{align}
The relations
\begin{align}
	&\int \frac{d\bm k}{(2\pi)^2}\ \sinh(\frac{\bm{q'}}{2}\cdot \frac{\partial}{\partial \bm k}) W^{s} (\bm{q},\bm{k},t) = 0 
\end{align}
and 
\begin{align}
	&\int \frac{d\bm k}{(2\pi)^2}\ \bm k \sinh(\frac{\bm{q'}}{2}\cdot \frac{\partial}{\partial \bm k}) W^{s} (\bm{q},\bm{k},t) = -\frac{\bm q'}{2} n^s(\bm q,t)
\end{align}
follow after integration by parts and further assuming that the Wigner elements and all their derivatives with respect to $k_i$ approach zero as $k_i \rightarrow \pm \infty$. By further replacing $ n^s(\bm q, t)$ by $\int d\bm r' \ e^{-i\bm q' \cdot \bm r'} n^s(\bm r',t) $ and using the identities $e^{-\bm{q'}\cdot \frac{\partial}{\partial \bm q}} e^{-i\bm q\cdot \bm r'} = e^{i(\bm q' - \bm q)\bm r'}$, $i\bm q e^{i\bm q\cdot \bm r} = (\partial/\partial \bm r) e^{i\bm q\cdot \bm r}$, and $\int d\bm q \ e^{i(\bm r - \bm r')\bm q} (2\pi)^{-2} = \delta(\bm r - \bm r')$, we arrive at	
\begin{equation}
\frac{\partial n^{s} }{\partial t} = -\frac{2}{\hbar}\int \frac{d\bm k d\bm q}{(2\pi)^4} \ ie^{i\bm{r}\cdot\bm{q}} \ \Bigg[\sinh(\frac{\bm{q}}{2}\cdot \frac{\partial}{\partial \bm k})\mathcal E_{\bm k}^s\Bigg] \ W^{s}( \bm{q}, \bm{k}, t) , \label{nn1}
\end{equation}
and
\begin{equation}
\frac{\partial }{\partial t} (n^s\ov{\bm{p}}^{s}) = \ -2 \int \frac{d\bm k d\bm q}{(2\pi)^4}\ ie^{i\bm{r}\cdot\bm{q}} \ \bm{p} \ \Bigg[\sinh(\frac{\bm{q}}{2}\cdot \frac{\partial}{\partial \bm k})\mathcal E_{\bm k}^s\Bigg] \ W^{s}( \bm{q}, \bm{k}, t) - n^s\frac{\partial V}{\partial \bm r} . \label{pp1}
\end{equation}
In order to clarify the meaning of some of the terms in Eqs.~\eqref{nn1} and \eqref{pp1}, it is instructive to derive the classical limit $\hbar \rightarrow 0$, which must be taken after replacing $\hbar \bm k$ by $\bm p$ and keeping it finite. Performing the Taylor expansion of the $\sinh(\cdot)$ operator in Eqs.~\eqref{nn1} and \eqref{pp1} and some algebra, we are led to Eqs.~\eqref{nnSep1} and \eqref{kkSep1}.

\section{Fluid variables for $\xi \ll 1$}
\label{apB}

Here we give explicit expressions for the linearised fluid quantities, which can all be expanded in even powers of the reduced velocity $\xi(\bm r,t)$. For the $T\rightarrow  0$ limit, we get
\begin{align}
	&n  \simeq n_0\Big(1 + \frac{3}{2}\xi^2\Big), \label{apn1}\\
	&m  \simeq \mathcal M \Big( 1+ \xi^2\Big),\\
	&[P_{\rm cl}]_{xx}   \simeq \frac{p_{\rm F}v_{\rm F}n_0}{3\pi^2} \Big(1+\frac{9}{2}\xi^2\Big),\\
	&\ov{J_{xxx}}  \simeq 0,\\
	&\ov{T_{xxx}^x}  \simeq -\frac{3}{4\pi^2} \frac{n_0}{\mathcal Mn} \Big(1 + \frac{1}{2}\xi^2\Big).\label{apn4}
\end{align}
On the contrary, the momentum along $x$ holds an expansion in odd powers of $\xi$,  
\begin{align}
	\ov{p}_x \simeq v_{\rm F}\mathcal M\Big( \xi + \xi^3\Big)
\end{align}
Plugging Eqs.~\eqref{apn1}-\eqref{apn4} into Eqs.~\eqref{finalnn} and \eqref{finalkk}, and neglecting the $\order{\xi^3}$ terms, yields Eqs.~\eqref{qc1} and \eqref{qc2}.
\end{appendices}
\twocolumngrid

\bibliography{manuscript_v3.bib}

\begin{thebibliography}{55}%
\makeatletter
\providecommand \@ifxundefined [1]{%
 \@ifx{#1\undefined}
}%
\providecommand \@ifnum [1]{%
 \ifnum #1\expandafter \@firstoftwo
 \else \expandafter \@secondoftwo
 \fi
}%
\providecommand \@ifx [1]{%
 \ifx #1\expandafter \@firstoftwo
 \else \expandafter \@secondoftwo
 \fi
}%
\providecommand \natexlab [1]{#1}%
\providecommand \enquote  [1]{``#1''}%
\providecommand \bibnamefont  [1]{#1}%
\providecommand \bibfnamefont [1]{#1}%
\providecommand \citenamefont [1]{#1}%
\providecommand \href@noop [0]{\@secondoftwo}%
\providecommand \href [0]{\begingroup \@sanitize@url \@href}%
\providecommand \@href[1]{\@@startlink{#1}\@@href}%
\providecommand \@@href[1]{\endgroup#1\@@endlink}%
\providecommand \@sanitize@url [0]{\catcode `\\12\catcode `\$12\catcode
  `\&12\catcode `\#12\catcode `\^12\catcode `\_12\catcode `\%12\relax}%
\providecommand \@@startlink[1]{}%
\providecommand \@@endlink[0]{}%
\providecommand \url  [0]{\begingroup\@sanitize@url \@url }%
\providecommand \@url [1]{\endgroup\@href {#1}{\urlprefix }}%
\providecommand \urlprefix  [0]{URL }%
\providecommand \Eprint [0]{\href }%
\providecommand \doibase [0]{http://dx.doi.org/}%
\providecommand \selectlanguage [0]{\@gobble}%
\providecommand \bibinfo  [0]{\@secondoftwo}%
\providecommand \bibfield  [0]{\@secondoftwo}%
\providecommand \translation [1]{[#1]}%
\providecommand \BibitemOpen [0]{}%
\providecommand \bibitemStop [0]{}%
\providecommand \bibitemNoStop [0]{.\EOS\space}%
\providecommand \EOS [0]{\spacefactor3000\relax}%
\providecommand \BibitemShut  [1]{\csname bibitem#1\endcsname}%
\let\auto@bib@innerbib\@empty
%</preamble>
\bibitem [{\citenamefont {Castro~Neto}\ \emph {et~al.}(2009)\citenamefont
  {Castro~Neto}, \citenamefont {Guinea}, \citenamefont {Peres}, \citenamefont
  {Novoselov},\ and\ \citenamefont {Geim}}]{Castroetall}%
  \BibitemOpen
  \bibfield  {author} {\bibinfo {author} {\bibfnamefont {A.~H.}\ \bibnamefont
  {Castro~Neto}}, \bibinfo {author} {\bibfnamefont {F.}~\bibnamefont {Guinea}},
  \bibinfo {author} {\bibfnamefont {N.~M.~R.}\ \bibnamefont {Peres}}, \bibinfo
  {author} {\bibfnamefont {K.~S.}\ \bibnamefont {Novoselov}}, \ and\ \bibinfo
  {author} {\bibfnamefont {A.~K.}\ \bibnamefont {Geim}},\ }\href {\doibase
  10.1103/RevModPhys.81.109} {\bibfield  {journal} {\bibinfo  {journal} {Rev.
  Mod. Phys.}\ }\textbf {\bibinfo {volume} {81}},\ \bibinfo {pages} {109}
  (\bibinfo {year} {2009})}\BibitemShut {NoStop}%
\bibitem [{\citenamefont {Wallace}(1947)}]{Wallace}%
  \BibitemOpen
  \bibfield  {author} {\bibinfo {author} {\bibfnamefont {P.~R.}\ \bibnamefont
  {Wallace}},\ }\href {\doibase 10.1103/PhysRev.71.622} {\bibfield  {journal}
  {\bibinfo  {journal} {Phys. Rev.}\ }\textbf {\bibinfo {volume} {71}},\
  \bibinfo {pages} {622} (\bibinfo {year} {1947})}\BibitemShut {NoStop}%
\bibitem [{\citenamefont {Rodrigo}\ \emph {et~al.}(2015)\citenamefont
  {Rodrigo}, \citenamefont {Limaj}, \citenamefont {Janner}, \citenamefont
  {Etezadi}, \citenamefont {Garcia~de Abajo}, \citenamefont {Pruneri},\ and\
  \citenamefont {Altug}}]{sensing1}%
  \BibitemOpen
  \bibfield  {author} {\bibinfo {author} {\bibfnamefont {D.}~\bibnamefont
  {Rodrigo}}, \bibinfo {author} {\bibfnamefont {O.}~\bibnamefont {Limaj}},
  \bibinfo {author} {\bibfnamefont {D.}~\bibnamefont {Janner}}, \bibinfo
  {author} {\bibfnamefont {D.}~\bibnamefont {Etezadi}}, \bibinfo {author}
  {\bibfnamefont {F.~J.}\ \bibnamefont {Garcia~de Abajo}}, \bibinfo {author}
  {\bibfnamefont {V.}~\bibnamefont {Pruneri}}, \ and\ \bibinfo {author}
  {\bibfnamefont {H.}~\bibnamefont {Altug}},\ }\href {\doibase
  10.1126/science.aab2051} {\bibfield  {journal} {\bibinfo  {journal}
  {Science}\ }\textbf {\bibinfo {volume} {349}},\ \bibinfo {pages} {165–168}
  (\bibinfo {year} {2015})}\BibitemShut {NoStop}%
\bibitem [{\citenamefont {Chen}\ \emph {et~al.}(2012)\citenamefont {Chen},
  \citenamefont {Badioli}, \citenamefont {Alonso-González}, \citenamefont
  {Thongrattanasiri}, \citenamefont {Huth}, \citenamefont {Osmond},
  \citenamefont {Spasenović}, \citenamefont {Centeno}, \citenamefont
  {Pesquera}, \citenamefont {Godignon},\ and\ \citenamefont
  {et~al.}}]{sensing2}%
  \BibitemOpen
  \bibfield  {author} {\bibinfo {author} {\bibfnamefont {J.}~\bibnamefont
  {Chen}}, \bibinfo {author} {\bibfnamefont {M.}~\bibnamefont {Badioli}},
  \bibinfo {author} {\bibfnamefont {P.}~\bibnamefont {Alonso-González}},
  \bibinfo {author} {\bibfnamefont {S.}~\bibnamefont {Thongrattanasiri}},
  \bibinfo {author} {\bibfnamefont {F.}~\bibnamefont {Huth}}, \bibinfo {author}
  {\bibfnamefont {J.}~\bibnamefont {Osmond}}, \bibinfo {author} {\bibfnamefont
  {M.}~\bibnamefont {Spasenović}}, \bibinfo {author} {\bibfnamefont
  {A.}~\bibnamefont {Centeno}}, \bibinfo {author} {\bibfnamefont
  {A.}~\bibnamefont {Pesquera}}, \bibinfo {author} {\bibfnamefont
  {P.}~\bibnamefont {Godignon}}, \ and\ \bibinfo {author} {\bibnamefont
  {et~al.}},\ }\href {\doibase 10.1038/nature11254} {\bibfield  {journal}
  {\bibinfo  {journal} {Nature}\ }\textbf {\bibinfo {volume} {487}},\ \bibinfo
  {pages} {77–81} (\bibinfo {year} {2012})}\BibitemShut {NoStop}%
\bibitem [{\citenamefont {Zeng}\ \emph {et~al.}(2015)\citenamefont {Zeng},
  \citenamefont {Sreekanth}, \citenamefont {Shang}, \citenamefont {Yu},
  \citenamefont {Chen}, \citenamefont {Yin}, \citenamefont {Baillargeat},
  \citenamefont {Coquet}, \citenamefont {Ho}, \citenamefont {Kabashin},\ and\
  \citenamefont {Yong}}]{sensing3}%
  \BibitemOpen
  \bibfield  {author} {\bibinfo {author} {\bibfnamefont {S.}~\bibnamefont
  {Zeng}}, \bibinfo {author} {\bibfnamefont {K.~V.}\ \bibnamefont {Sreekanth}},
  \bibinfo {author} {\bibfnamefont {J.}~\bibnamefont {Shang}}, \bibinfo
  {author} {\bibfnamefont {T.}~\bibnamefont {Yu}}, \bibinfo {author}
  {\bibfnamefont {C.-K.}\ \bibnamefont {Chen}}, \bibinfo {author}
  {\bibfnamefont {F.}~\bibnamefont {Yin}}, \bibinfo {author} {\bibfnamefont
  {D.}~\bibnamefont {Baillargeat}}, \bibinfo {author} {\bibfnamefont
  {P.}~\bibnamefont {Coquet}}, \bibinfo {author} {\bibfnamefont {H.-P.}\
  \bibnamefont {Ho}}, \bibinfo {author} {\bibfnamefont {A.~V.}\ \bibnamefont
  {Kabashin}}, \ and\ \bibinfo {author} {\bibfnamefont {K.-T.}\ \bibnamefont
  {Yong}},\ }\href {\doibase 10.1002/adma.201501754} {\bibfield  {journal}
  {\bibinfo  {journal} {Advanced Materials}\ }\textbf {\bibinfo {volume}
  {27}},\ \bibinfo {pages} {6163} (\bibinfo {year} {2015})}\BibitemShut
  {NoStop}%
\bibitem [{\citenamefont {Koppens}\ \emph {et~al.}(2011)\citenamefont
  {Koppens}, \citenamefont {Chang},\ and\ \citenamefont {García~de
  Abajo}}]{Koppens}%
  \BibitemOpen
  \bibfield  {author} {\bibinfo {author} {\bibfnamefont {F.~H.~L.}\
  \bibnamefont {Koppens}}, \bibinfo {author} {\bibfnamefont {D.~E.}\
  \bibnamefont {Chang}}, \ and\ \bibinfo {author} {\bibfnamefont {F.~J.}\
  \bibnamefont {García~de Abajo}},\ }\href {\doibase 10.1021/nl201771h}
  {\bibfield  {journal} {\bibinfo  {journal} {Nano Letters}\ }\textbf {\bibinfo
  {volume} {11}},\ \bibinfo {pages} {3370} (\bibinfo {year} {2011})},\ \bibinfo
  {note} {pMID: 21766812}\BibitemShut {NoStop}%
\bibitem [{\citenamefont {Agarwal}\ \emph {et~al.}(2018)\citenamefont
  {Agarwal}, \citenamefont {Vitiello}, \citenamefont {Viti}, \citenamefont
  {Cupolillo},\ and\ \citenamefont {Politano}}]{plasmonics1}%
  \BibitemOpen
  \bibfield  {author} {\bibinfo {author} {\bibfnamefont {A.}~\bibnamefont
  {Agarwal}}, \bibinfo {author} {\bibfnamefont {M.}~\bibnamefont {Vitiello}},
  \bibinfo {author} {\bibfnamefont {L.}~\bibnamefont {Viti}}, \bibinfo {author}
  {\bibfnamefont {A.}~\bibnamefont {Cupolillo}}, \ and\ \bibinfo {author}
  {\bibfnamefont {A.}~\bibnamefont {Politano}},\ }\href {\doibase
  10.1039/C8NR01395K} {\bibfield  {journal} {\bibinfo  {journal} {Nanoscale}\
  }\textbf {\bibinfo {volume} {10}} (\bibinfo {year} {2018}),\
  10.1039/C8NR01395K}\BibitemShut {NoStop}%
\bibitem [{\citenamefont {Dai}\ \emph {et~al.}(2014)\citenamefont {Dai},
  \citenamefont {Fei}, \citenamefont {Ma}, \citenamefont {Rodin}, \citenamefont
  {Wagner}, \citenamefont {McLeod}, \citenamefont {Liu}, \citenamefont
  {Gannett}, \citenamefont {Regan}, \citenamefont {Watanabe}, \citenamefont
  {Taniguchi}, \citenamefont {Thiemens}, \citenamefont {Dominguez},
  \citenamefont {Neto}, \citenamefont {Zettl}, \citenamefont {Keilmann},
  \citenamefont {Jarillo-Herrero}, \citenamefont {Fogler},\ and\ \citenamefont
  {Basov}}]{Dai1125}%
  \BibitemOpen
  \bibfield  {author} {\bibinfo {author} {\bibfnamefont {S.}~\bibnamefont
  {Dai}}, \bibinfo {author} {\bibfnamefont {Z.}~\bibnamefont {Fei}}, \bibinfo
  {author} {\bibfnamefont {Q.}~\bibnamefont {Ma}}, \bibinfo {author}
  {\bibfnamefont {A.~S.}\ \bibnamefont {Rodin}}, \bibinfo {author}
  {\bibfnamefont {M.}~\bibnamefont {Wagner}}, \bibinfo {author} {\bibfnamefont
  {A.~S.}\ \bibnamefont {McLeod}}, \bibinfo {author} {\bibfnamefont {M.~K.}\
  \bibnamefont {Liu}}, \bibinfo {author} {\bibfnamefont {W.}~\bibnamefont
  {Gannett}}, \bibinfo {author} {\bibfnamefont {W.}~\bibnamefont {Regan}},
  \bibinfo {author} {\bibfnamefont {K.}~\bibnamefont {Watanabe}}, \bibinfo
  {author} {\bibfnamefont {T.}~\bibnamefont {Taniguchi}}, \bibinfo {author}
  {\bibfnamefont {M.}~\bibnamefont {Thiemens}}, \bibinfo {author}
  {\bibfnamefont {G.}~\bibnamefont {Dominguez}}, \bibinfo {author}
  {\bibfnamefont {A.~H.~C.}\ \bibnamefont {Neto}}, \bibinfo {author}
  {\bibfnamefont {A.}~\bibnamefont {Zettl}}, \bibinfo {author} {\bibfnamefont
  {F.}~\bibnamefont {Keilmann}}, \bibinfo {author} {\bibfnamefont
  {P.}~\bibnamefont {Jarillo-Herrero}}, \bibinfo {author} {\bibfnamefont
  {M.~M.}\ \bibnamefont {Fogler}}, \ and\ \bibinfo {author} {\bibfnamefont
  {D.~N.}\ \bibnamefont {Basov}},\ }\href {\doibase 10.1126/science.1246833}
  {\bibfield  {journal} {\bibinfo  {journal} {Science}\ }\textbf {\bibinfo
  {volume} {343}},\ \bibinfo {pages} {1125} (\bibinfo {year}
  {2014})}\BibitemShut {NoStop}%
\bibitem [{\citenamefont {Lin}\ \emph {et~al.}(2009)\citenamefont {Lin},
  \citenamefont {Williams},\ and\ \citenamefont {Connell}}]{Lin2009}%
  \BibitemOpen
  \bibfield  {author} {\bibinfo {author} {\bibfnamefont {Y.}~\bibnamefont
  {Lin}}, \bibinfo {author} {\bibfnamefont {T.~V.}\ \bibnamefont {Williams}}, \
  and\ \bibinfo {author} {\bibfnamefont {J.~W.}\ \bibnamefont {Connell}},\
  }\href {\doibase 10.1021/jz9002108} {\bibfield  {journal} {\bibinfo
  {journal} {The Journal of Physical Chemistry Letters}\ }\textbf {\bibinfo
  {volume} {1}},\ \bibinfo {pages} {277} (\bibinfo {year} {2009})}\BibitemShut
  {NoStop}%
\bibitem [{\citenamefont {Li}\ \emph {et~al.}(2017)\citenamefont {Li},
  \citenamefont {Li}, \citenamefont {Chi}, \citenamefont {Shan}, \citenamefont
  {Zheng},\ and\ \citenamefont {Fang}}]{plasmonics2}%
  \BibitemOpen
  \bibfield  {author} {\bibinfo {author} {\bibfnamefont {Y.}~\bibnamefont
  {Li}}, \bibinfo {author} {\bibfnamefont {Z.}~\bibnamefont {Li}}, \bibinfo
  {author} {\bibfnamefont {C.}~\bibnamefont {Chi}}, \bibinfo {author}
  {\bibfnamefont {H.}~\bibnamefont {Shan}}, \bibinfo {author} {\bibfnamefont
  {L.}~\bibnamefont {Zheng}}, \ and\ \bibinfo {author} {\bibfnamefont
  {Z.}~\bibnamefont {Fang}},\ }\href {\doibase 10.1002/advs.201600430}
  {\bibfield  {journal} {\bibinfo  {journal} {Advanced Science}\ }\textbf
  {\bibinfo {volume} {4}},\ \bibinfo {pages} {1600430} (\bibinfo {year}
  {2017})}\BibitemShut {NoStop}%
\bibitem [{\citenamefont {Grigorenko}\ \emph {et~al.}(2012)\citenamefont
  {Grigorenko}, \citenamefont {Polini},\ and\ \citenamefont
  {Novoselov}}]{Grigorenko2012}%
  \BibitemOpen
  \bibfield  {author} {\bibinfo {author} {\bibfnamefont {A.~N.}\ \bibnamefont
  {Grigorenko}}, \bibinfo {author} {\bibfnamefont {M.}~\bibnamefont {Polini}},
  \ and\ \bibinfo {author} {\bibfnamefont {K.~S.}\ \bibnamefont {Novoselov}},\
  }\href {\doibase 10.1038/nphoton.2012.262} {\bibfield  {journal} {\bibinfo
  {journal} {Nature Photonics}\ }\textbf {\bibinfo {volume} {6}},\ \bibinfo
  {pages} {749} (\bibinfo {year} {2012})}\BibitemShut {NoStop}%
\bibitem [{\citenamefont {Geim}\ and\ \citenamefont {Novoselov}(2007)}]{Geim1}%
  \BibitemOpen
  \bibfield  {author} {\bibinfo {author} {\bibfnamefont {A.}~\bibnamefont
  {Geim}}\ and\ \bibinfo {author} {\bibfnamefont {K.}~\bibnamefont
  {Novoselov}},\ }\href {\doibase 10.1038/nmat1849} {\bibfield  {journal}
  {\bibinfo  {journal} {Nature materials}\ }\textbf {\bibinfo {volume} {6}},\
  \bibinfo {pages} {183} (\bibinfo {year} {2007})}\BibitemShut {NoStop}%
\bibitem [{\citenamefont {Yang}\ \emph {et~al.}(2016)\citenamefont {Yang},
  \citenamefont {Shi}, \citenamefont {Li},\ and\ \citenamefont {Li}}]{gating}%
  \BibitemOpen
  \bibfield  {author} {\bibinfo {author} {\bibfnamefont {Y.}~\bibnamefont
  {Yang}}, \bibinfo {author} {\bibfnamefont {Z.}~\bibnamefont {Shi}}, \bibinfo
  {author} {\bibfnamefont {J.}~\bibnamefont {Li}}, \ and\ \bibinfo {author}
  {\bibfnamefont {Z.-Y.}\ \bibnamefont {Li}},\ }\href {\doibase
  10.1364/PRJ.4.000065} {\bibfield  {journal} {\bibinfo  {journal} {Photonics
  Research}\ }\textbf {\bibinfo {volume} {4}},\ \bibinfo {pages} {65} (\bibinfo
  {year} {2016})}\BibitemShut {NoStop}%
\bibitem [{\citenamefont {Wang}\ \emph {et~al.}(2008)\citenamefont {Wang},
  \citenamefont {Zhang}, \citenamefont {Tian}, \citenamefont {Girit},
  \citenamefont {Zettl}, \citenamefont {Crommie},\ and\ \citenamefont
  {Shen}}]{Wang}%
  \BibitemOpen
  \bibfield  {author} {\bibinfo {author} {\bibfnamefont {F.}~\bibnamefont
  {Wang}}, \bibinfo {author} {\bibfnamefont {Y.}~\bibnamefont {Zhang}},
  \bibinfo {author} {\bibfnamefont {C.}~\bibnamefont {Tian}}, \bibinfo {author}
  {\bibfnamefont {C.}~\bibnamefont {Girit}}, \bibinfo {author} {\bibfnamefont
  {A.}~\bibnamefont {Zettl}}, \bibinfo {author} {\bibfnamefont
  {M.}~\bibnamefont {Crommie}}, \ and\ \bibinfo {author} {\bibfnamefont
  {Y.~R.}\ \bibnamefont {Shen}},\ }\href {\doibase 10.1126/science.1152793}
  {\bibfield  {journal} {\bibinfo  {journal} {Science}\ }\textbf {\bibinfo
  {volume} {320}},\ \bibinfo {pages} {206} (\bibinfo {year}
  {2008})}\BibitemShut {NoStop}%
\bibitem [{\citenamefont {Laboratory}\ \emph {et~al.}(2008)\citenamefont
  {Laboratory}, \citenamefont {of~Energy. Office~of Scientific},\ and\
  \citenamefont {Information}}]{lawrence}%
  \BibitemOpen
  \bibfield  {author} {\bibinfo {author} {\bibfnamefont {L.~B.~N.}\
  \bibnamefont {Laboratory}}, \bibinfo {author} {\bibfnamefont {U.~S.~D.}\
  \bibnamefont {of~Energy. Office~of Scientific}}, \ and\ \bibinfo {author}
  {\bibfnamefont {T.}~\bibnamefont {Information}},\ }\href
  {https://books.google.pt/books?id=G8AtngAACAAJ} {\emph {\bibinfo {title}
  {Dirac Charge Dynamics in Graphene by Infrared Spectroscopy}}}\ (\bibinfo
  {publisher} {Lawrence Berkeley National Laboratory},\ \bibinfo {year}
  {2008})\BibitemShut {NoStop}%
\bibitem [{\citenamefont {Cosme}\ and\ \citenamefont
  {Ter\c{c}as}(2020)}]{graf4}%
  \BibitemOpen
  \bibfield  {author} {\bibinfo {author} {\bibfnamefont {P.}~\bibnamefont
  {Cosme}}\ and\ \bibinfo {author} {\bibfnamefont {H.}~\bibnamefont
  {Ter\c{c}as}},\ }\href {\doibase 10.1021/acsphotonics.0c00313} {\bibfield
  {journal} {\bibinfo  {journal} {ACS Photonics}\ }\textbf {\bibinfo {volume}
  {7}},\ \bibinfo {pages} {1375} (\bibinfo {year} {2020})}\BibitemShut
  {NoStop}%
\bibitem [{\citenamefont {Ryzhii}\ \emph {et~al.}(2007)\citenamefont {Ryzhii},
  \citenamefont {Satou},\ and\ \citenamefont {Otsuji}}]{graf1}%
  \BibitemOpen
  \bibfield  {author} {\bibinfo {author} {\bibfnamefont {V.}~\bibnamefont
  {Ryzhii}}, \bibinfo {author} {\bibfnamefont {A.}~\bibnamefont {Satou}}, \
  and\ \bibinfo {author} {\bibfnamefont {T.}~\bibnamefont {Otsuji}},\ }\href
  {\doibase 10.1063/1.2426904} {\bibfield  {journal} {\bibinfo  {journal}
  {Journal of Applied Physics}\ }\textbf {\bibinfo {volume} {101}} (\bibinfo
  {year} {2007}),\ 10.1063/1.2426904}\BibitemShut {NoStop}%
\bibitem [{\citenamefont {Bistritzer}\ and\ \citenamefont
  {MacDonald}(2009)}]{graf2}%
  \BibitemOpen
  \bibfield  {author} {\bibinfo {author} {\bibfnamefont {R.}~\bibnamefont
  {Bistritzer}}\ and\ \bibinfo {author} {\bibfnamefont {A.~H.}\ \bibnamefont
  {MacDonald}},\ }\href {\doibase 10.1103/PhysRevB.80.085109} {\bibfield
  {journal} {\bibinfo  {journal} {Phys. Rev. B}\ }\textbf {\bibinfo {volume}
  {80}},\ \bibinfo {pages} {085109} (\bibinfo {year} {2009})}\BibitemShut
  {NoStop}%
\bibitem [{\citenamefont {Svintsov}\ \emph {et~al.}(2012)\citenamefont
  {Svintsov}, \citenamefont {Vyurkov}, \citenamefont {Yurchenko}, \citenamefont
  {Otsuji},\ and\ \citenamefont {Ryzhii}}]{graf3}%
  \BibitemOpen
  \bibfield  {author} {\bibinfo {author} {\bibfnamefont {D.}~\bibnamefont
  {Svintsov}}, \bibinfo {author} {\bibfnamefont {V.}~\bibnamefont {Vyurkov}},
  \bibinfo {author} {\bibfnamefont {S.}~\bibnamefont {Yurchenko}}, \bibinfo
  {author} {\bibfnamefont {T.}~\bibnamefont {Otsuji}}, \ and\ \bibinfo {author}
  {\bibfnamefont {V.}~\bibnamefont {Ryzhii}},\ }\href {\doibase
  10.1063/1.4705382} {\bibfield  {journal} {\bibinfo  {journal} {Journal of
  Applied Physics}\ }\textbf {\bibinfo {volume} {111}},\ \bibinfo {pages}
  {083715} (\bibinfo {year} {2012})}\BibitemShut {NoStop}%
\bibitem [{\citenamefont {Schwengelbeck}\ \emph {et~al.}(2000)\citenamefont
  {Schwengelbeck}, \citenamefont {Plaja}, \citenamefont {Roso},\ and\
  \citenamefont {Jarque}}]{HFintroduction1}%
  \BibitemOpen
  \bibfield  {author} {\bibinfo {author} {\bibfnamefont {U.}~\bibnamefont
  {Schwengelbeck}}, \bibinfo {author} {\bibfnamefont {L.}~\bibnamefont
  {Plaja}}, \bibinfo {author} {\bibfnamefont {L.}~\bibnamefont {Roso}}, \ and\
  \bibinfo {author} {\bibfnamefont {E.~C.}\ \bibnamefont {Jarque}},\ }\href
  {\doibase 10.1088/0953-4075/33/8/314} {\bibfield  {journal} {\bibinfo
  {journal} {Journal of Physics B: Atomic, Molecular and Optical Physics}\
  }\textbf {\bibinfo {volume} {33}},\ \bibinfo {pages} {1653} (\bibinfo {year}
  {2000})}\BibitemShut {NoStop}%
\bibitem [{\citenamefont {Jasiak}\ \emph {et~al.}(2009)\citenamefont {Jasiak},
  \citenamefont {Manfredi}, \citenamefont {Hervieux},\ and\ \citenamefont
  {Haefele}}]{HFintroduction2}%
  \BibitemOpen
  \bibfield  {author} {\bibinfo {author} {\bibfnamefont {R.}~\bibnamefont
  {Jasiak}}, \bibinfo {author} {\bibfnamefont {G.}~\bibnamefont {Manfredi}},
  \bibinfo {author} {\bibfnamefont {P.-A.}\ \bibnamefont {Hervieux}}, \ and\
  \bibinfo {author} {\bibfnamefont {M.}~\bibnamefont {Haefele}},\ }\href
  {\doibase 10.1088/1367-2630/11/6/063042} {\bibfield  {journal} {\bibinfo
  {journal} {New Journal of Physics}\ }\textbf {\bibinfo {volume} {11}},\
  \bibinfo {pages} {063042} (\bibinfo {year} {2009})}\BibitemShut {NoStop}%
\bibitem [{\citenamefont {Teperik}\ \emph {et~al.}(2013)\citenamefont
  {Teperik}, \citenamefont {Nordlander}, \citenamefont {Aizpurua},\ and\
  \citenamefont {Borisov}}]{DFTintroduction}%
  \BibitemOpen
  \bibfield  {author} {\bibinfo {author} {\bibfnamefont {T.~V.}\ \bibnamefont
  {Teperik}}, \bibinfo {author} {\bibfnamefont {P.}~\bibnamefont {Nordlander}},
  \bibinfo {author} {\bibfnamefont {J.}~\bibnamefont {Aizpurua}}, \ and\
  \bibinfo {author} {\bibfnamefont {A.~G.}\ \bibnamefont {Borisov}},\ }\href
  {\doibase 10.1103/PhysRevLett.110.263901} {\bibfield  {journal} {\bibinfo
  {journal} {Phys. Rev. Lett.}\ }\textbf {\bibinfo {volume} {110}},\ \bibinfo
  {pages} {263901} (\bibinfo {year} {2013})}\BibitemShut {NoStop}%
\bibitem [{\citenamefont {Dugaev}\ and\ \citenamefont
  {Katsnelson}(2013)}]{dugaev_2013}%
  \BibitemOpen
  \bibfield  {author} {\bibinfo {author} {\bibfnamefont {V.~K.}\ \bibnamefont
  {Dugaev}}\ and\ \bibinfo {author} {\bibfnamefont {M.~I.}\ \bibnamefont
  {Katsnelson}},\ }\href {\doibase 10.1103/PhysRevB.88.235432} {\bibfield
  {journal} {\bibinfo  {journal} {Phys. Rev. B}\ }\textbf {\bibinfo {volume}
  {88}},\ \bibinfo {pages} {235432} (\bibinfo {year} {2013})}\BibitemShut
  {NoStop}%
\bibitem [{\citenamefont {Wigner}(1932)}]{wigner}%
  \BibitemOpen
  \bibfield  {author} {\bibinfo {author} {\bibfnamefont {E.}~\bibnamefont
  {Wigner}},\ }\href {\doibase 10.1103/PhysRev.40.749} {\bibfield  {journal}
  {\bibinfo  {journal} {Phys. Rev.}\ }\textbf {\bibinfo {volume} {40}},\
  \bibinfo {pages} {749} (\bibinfo {year} {1932})}\BibitemShut {NoStop}%
\bibitem [{\citenamefont {Morandi}\ and\ \citenamefont
  {Schuerrer}(2011)}]{ans1}%
  \BibitemOpen
  \bibfield  {author} {\bibinfo {author} {\bibfnamefont {O.}~\bibnamefont
  {Morandi}}\ and\ \bibinfo {author} {\bibfnamefont {F.}~\bibnamefont
  {Schuerrer}},\ }\href {\doibase 10.1088/1751-8113/44/26/265301} {\bibfield
  {journal} {\bibinfo  {journal} {Journal of Physics A-mathematical and
  Theoretical - J PHYS A-MATH THEOR}\ }\textbf {\bibinfo {volume} {44}}
  (\bibinfo {year} {2011}),\ 10.1088/1751-8113/44/26/265301}\BibitemShut
  {NoStop}%
\bibitem [{\citenamefont {Zamponi}\ and\ \citenamefont
  {Barletti}(2011)}]{ans2}%
  \BibitemOpen
  \bibfield  {author} {\bibinfo {author} {\bibfnamefont {N.}~\bibnamefont
  {Zamponi}}\ and\ \bibinfo {author} {\bibfnamefont {L.}~\bibnamefont
  {Barletti}},\ }\href {\doibase https://doi.org/10.1002/mma.1403} {\bibfield
  {journal} {\bibinfo  {journal} {Mathematical Methods in the Applied
  Sciences}\ }\textbf {\bibinfo {volume} {34}},\ \bibinfo {pages} {807}
  (\bibinfo {year} {2011})}\BibitemShut {NoStop}%
\bibitem [{\citenamefont {Robertson}(1966)}]{Liouville}%
  \BibitemOpen
  \bibfield  {author} {\bibinfo {author} {\bibfnamefont {B.}~\bibnamefont
  {Robertson}},\ }\href {\doibase 10.1103/PhysRev.144.151} {\bibfield
  {journal} {\bibinfo  {journal} {Phys. Rev.}\ }\textbf {\bibinfo {volume}
  {144}},\ \bibinfo {pages} {151} (\bibinfo {year} {1966})}\BibitemShut
  {NoStop}%
\bibitem [{\citenamefont {Weyl}(1927)}]{weyl}%
  \BibitemOpen
  \bibfield  {author} {\bibinfo {author} {\bibfnamefont {H.}~\bibnamefont
  {Weyl}},\ }\href@noop {} {\bibfield  {journal} {\bibinfo  {journal} {Z.
  Phys.}\ }\textbf {\bibinfo {volume} {46}} (\bibinfo {year}
  {1927})}\BibitemShut {NoStop}%
\bibitem [{\citenamefont {Moyal}(1949)}]{moyal}%
  \BibitemOpen
  \bibfield  {author} {\bibinfo {author} {\bibfnamefont {J.~E.}\ \bibnamefont
  {Moyal}},\ }\href@noop {} {\bibfield  {journal} {\bibinfo  {journal} {Proc.
  Cambridge Phil. Soc.}\ }\textbf {\bibinfo {volume} {45}},\ \bibinfo {pages}
  {99} (\bibinfo {year} {1949})}\BibitemShut {NoStop}%
\bibitem [{\citenamefont {Hillery}\ \emph {et~al.}(1984)\citenamefont
  {Hillery}, \citenamefont {O'Connell}, \citenamefont {Scully},\ and\
  \citenamefont {Wigner}}]{hillery}%
  \BibitemOpen
  \bibfield  {author} {\bibinfo {author} {\bibfnamefont {M.}~\bibnamefont
  {Hillery}}, \bibinfo {author} {\bibfnamefont {R.}~\bibnamefont {O'Connell}},
  \bibinfo {author} {\bibfnamefont {M.}~\bibnamefont {Scully}}, \ and\ \bibinfo
  {author} {\bibfnamefont {E.}~\bibnamefont {Wigner}},\ }\href {\doibase
  https://doi.org/10.1016/0370-1573(84)90160-1} {\bibfield  {journal} {\bibinfo
   {journal} {Physics Reports}\ }\textbf {\bibinfo {volume} {106}},\ \bibinfo
  {pages} {121 } (\bibinfo {year} {1984})}\BibitemShut {NoStop}%
\bibitem [{\citenamefont {Groot}\ \emph {et~al.}(1974)\citenamefont {Groot},
  \citenamefont {Weyl},\ and\ \citenamefont {Wigner}}]{groot}%
  \BibitemOpen
  \bibfield  {author} {\bibinfo {author} {\bibfnamefont {S.}~\bibnamefont
  {Groot}}, \bibinfo {author} {\bibfnamefont {H.}~\bibnamefont {Weyl}}, \ and\
  \bibinfo {author} {\bibfnamefont {E.}~\bibnamefont {Wigner}},\ }\href
  {https://books.google.pt/books?id=FJQuAAAAIAAJ} {\emph {\bibinfo {title} {La
  transformation de Weyl et la fonction de Wigner: une forme alternative de la
  m{\'e}canique quantique}}},\ Collection de la Chaire Aisenstadt\ (\bibinfo
  {publisher} {Les Presses de l'Universiti{\'e} de Montr{\'e}al},\ \bibinfo
  {year} {1974})\BibitemShut {NoStop}%
\bibitem [{\citenamefont {Das~Sarma}\ \emph {et~al.}(2009)\citenamefont
  {Das~Sarma}, \citenamefont {Hwang},\ and\ \citenamefont {Li}}]{pseudospin}%
  \BibitemOpen
  \bibfield  {author} {\bibinfo {author} {\bibfnamefont {S.}~\bibnamefont
  {Das~Sarma}}, \bibinfo {author} {\bibfnamefont {E.}~\bibnamefont {Hwang}}, \
  and\ \bibinfo {author} {\bibfnamefont {Q.}~\bibnamefont {Li}},\ }\href
  {\doibase 10.1103/PhysRevB.80.121303} {\bibfield  {journal} {\bibinfo
  {journal} {Phys. Rev. B}\ }\textbf {\bibinfo {volume} {80}} (\bibinfo {year}
  {2009}),\ 10.1103/PhysRevB.80.121303}\BibitemShut {NoStop}%
\bibitem [{\citenamefont {Fetter}\ and\ \citenamefont
  {Walecka}(1971)}]{walecka}%
  \BibitemOpen
  \bibfield  {author} {\bibinfo {author} {\bibfnamefont {A.~L.}\ \bibnamefont
  {Fetter}}\ and\ \bibinfo {author} {\bibfnamefont {J.~D.}\ \bibnamefont
  {Walecka}},\ }\href@noop {} {\emph {\bibinfo {title} {Quantum Theory of
  Many-Particle Systems}}}\ (\bibinfo  {publisher} {McGraw-Hill},\ \bibinfo
  {address} {Boston},\ \bibinfo {year} {1971})\BibitemShut {NoStop}%
\bibitem [{\citenamefont {Das~Sarma}\ and\ \citenamefont
  {Hwang}(2009)}]{DiracPlasma}%
  \BibitemOpen
  \bibfield  {author} {\bibinfo {author} {\bibfnamefont {S.}~\bibnamefont
  {Das~Sarma}}\ and\ \bibinfo {author} {\bibfnamefont {E.~H.}\ \bibnamefont
  {Hwang}},\ }\href {\doibase 10.1103/physrevlett.102.206412} {\bibfield
  {journal} {\bibinfo  {journal} {Physical Review Letters}\ }\textbf {\bibinfo
  {volume} {102}} (\bibinfo {year} {2009}),\
  10.1103/physrevlett.102.206412}\BibitemShut {NoStop}%
\bibitem [{\citenamefont {Wunsch}\ \emph
  {et~al.}(2006{\natexlab{a}})\citenamefont {Wunsch}, \citenamefont {Stauber},
  \citenamefont {Sols},\ and\ \citenamefont {Guinea}}]{Wunsch_2006}%
  \BibitemOpen
  \bibfield  {author} {\bibinfo {author} {\bibfnamefont {B.}~\bibnamefont
  {Wunsch}}, \bibinfo {author} {\bibfnamefont {T.}~\bibnamefont {Stauber}},
  \bibinfo {author} {\bibfnamefont {F.}~\bibnamefont {Sols}}, \ and\ \bibinfo
  {author} {\bibfnamefont {F.}~\bibnamefont {Guinea}},\ }\href {\doibase
  10.1088/1367-2630/8/12/318} {\bibfield  {journal} {\bibinfo  {journal} {New
  Journal of Physics}\ }\textbf {\bibinfo {volume} {8}},\ \bibinfo {pages}
  {318} (\bibinfo {year} {2006}{\natexlab{a}})}\BibitemShut {NoStop}%
\bibitem [{\citenamefont {Hwang}\ and\ \citenamefont
  {Das~Sarma}(2007)}]{Sarma2}%
  \BibitemOpen
  \bibfield  {author} {\bibinfo {author} {\bibfnamefont {E.~H.}\ \bibnamefont
  {Hwang}}\ and\ \bibinfo {author} {\bibfnamefont {S.}~\bibnamefont
  {Das~Sarma}},\ }\href {\doibase 10.1103/PhysRevB.75.205418} {\bibfield
  {journal} {\bibinfo  {journal} {Phys. Rev. B}\ }\textbf {\bibinfo {volume}
  {75}},\ \bibinfo {pages} {205418} (\bibinfo {year} {2007})}\BibitemShut
  {NoStop}%
\bibitem [{\citenamefont {Adam}\ \emph {et~al.}(2007)\citenamefont {Adam},
  \citenamefont {Hwang}, \citenamefont {Galitski},\ and\ \citenamefont
  {Das~Sarma}}]{Adam}%
  \BibitemOpen
  \bibfield  {author} {\bibinfo {author} {\bibfnamefont {S.}~\bibnamefont
  {Adam}}, \bibinfo {author} {\bibfnamefont {E.~H.}\ \bibnamefont {Hwang}},
  \bibinfo {author} {\bibfnamefont {V.~M.}\ \bibnamefont {Galitski}}, \ and\
  \bibinfo {author} {\bibfnamefont {S.}~\bibnamefont {Das~Sarma}},\ }\href
  {\doibase 10.1073/pnas.0704772104} {\bibfield  {journal} {\bibinfo  {journal}
  {Proceedings of the National Academy of Sciences}\ }\textbf {\bibinfo
  {volume} {104}},\ \bibinfo {pages} {18392–18397} (\bibinfo {year}
  {2007})}\BibitemShut {NoStop}%
\bibitem [{\citenamefont {Kittel}(1963)}]{kittel}%
  \BibitemOpen
  \bibfield  {author} {\bibinfo {author} {\bibfnamefont {C.}~\bibnamefont
  {Kittel}},\ }\href@noop {} {\emph {\bibinfo {title} {Quantum Theory of
  Solids}}}\ (\bibinfo  {publisher} {John Wiley and Sons, Inc},\ \bibinfo
  {year} {1963})\BibitemShut {NoStop}%
\bibitem [{\citenamefont {Wunsch}\ \emph
  {et~al.}(2006{\natexlab{b}})\citenamefont {Wunsch}, \citenamefont {Stauber},
  \citenamefont {Sols},\ and\ \citenamefont {Guinea}}]{finiteDoping}%
  \BibitemOpen
  \bibfield  {author} {\bibinfo {author} {\bibfnamefont {B.}~\bibnamefont
  {Wunsch}}, \bibinfo {author} {\bibfnamefont {T.}~\bibnamefont {Stauber}},
  \bibinfo {author} {\bibfnamefont {F.}~\bibnamefont {Sols}}, \ and\ \bibinfo
  {author} {\bibfnamefont {F.}~\bibnamefont {Guinea}},\ }\href {\doibase
  10.1088/1367-2630/8/12/318} {\bibfield  {journal} {\bibinfo  {journal} {New
  Journal of Physics}\ }\textbf {\bibinfo {volume} {8}},\ \bibinfo {pages}
  {318} (\bibinfo {year} {2006}{\natexlab{b}})}\BibitemShut {NoStop}%
\bibitem [{\citenamefont {Liu}\ \emph {et~al.}(2008)\citenamefont {Liu},
  \citenamefont {Willis}, \citenamefont {Emtsev},\ and\ \citenamefont
  {Seyller}}]{plasmon}%
  \BibitemOpen
  \bibfield  {author} {\bibinfo {author} {\bibfnamefont {Y.}~\bibnamefont
  {Liu}}, \bibinfo {author} {\bibfnamefont {R.}~\bibnamefont {Willis}},
  \bibinfo {author} {\bibfnamefont {K.}~\bibnamefont {Emtsev}}, \ and\ \bibinfo
  {author} {\bibfnamefont {T.}~\bibnamefont {Seyller}},\ }\href {\doibase
  10.1103/PhysRevB.78.201403} {\bibfield  {journal} {\bibinfo  {journal} {Phys.
  Rev. B}\ }\textbf {\bibinfo {volume} {78}} (\bibinfo {year} {2008}),\
  10.1103/PhysRevB.78.201403}\BibitemShut {NoStop}%
\bibitem [{\citenamefont {Zhu}\ \emph {et~al.}(2009)\citenamefont {Zhu},
  \citenamefont {Perebeinos}, \citenamefont {Freitag},\ and\ \citenamefont
  {Avouris}}]{capacitance}%
  \BibitemOpen
  \bibfield  {author} {\bibinfo {author} {\bibfnamefont {W.}~\bibnamefont
  {Zhu}}, \bibinfo {author} {\bibfnamefont {V.}~\bibnamefont {Perebeinos}},
  \bibinfo {author} {\bibfnamefont {M.}~\bibnamefont {Freitag}}, \ and\
  \bibinfo {author} {\bibfnamefont {P.}~\bibnamefont {Avouris}},\ }\href
  {\doibase 10.1103/PhysRevB.80.235402} {\bibfield  {journal} {\bibinfo
  {journal} {Phys. Rev. B}\ }\textbf {\bibinfo {volume} {80}},\ \bibinfo
  {pages} {235402} (\bibinfo {year} {2009})}\BibitemShut {NoStop}%
\bibitem [{\citenamefont {Stauber}\ \emph {et~al.}(2007)\citenamefont
  {Stauber}, \citenamefont {Peres},\ and\ \citenamefont
  {Guinea}}]{stauber_2007}%
  \BibitemOpen
  \bibfield  {author} {\bibinfo {author} {\bibfnamefont {T.}~\bibnamefont
  {Stauber}}, \bibinfo {author} {\bibfnamefont {N.~M.~R.}\ \bibnamefont
  {Peres}}, \ and\ \bibinfo {author} {\bibfnamefont {F.}~\bibnamefont
  {Guinea}},\ }\href {\doibase 10.1103/PhysRevB.76.205423} {\bibfield
  {journal} {\bibinfo  {journal} {Phys. Rev. B}\ }\textbf {\bibinfo {volume}
  {76}},\ \bibinfo {pages} {205423} (\bibinfo {year} {2007})}\BibitemShut
  {NoStop}%
\bibitem [{\citenamefont {Bizarro}\ \emph {et~al.}(2020)\citenamefont
  {Bizarro}, \citenamefont {Cortes},\ and\ \citenamefont
  {Vilela~Mendes}}]{bizarro_2020}%
  \BibitemOpen
  \bibfield  {author} {\bibinfo {author} {\bibfnamefont {J.~P.~S.}\
  \bibnamefont {Bizarro}}, \bibinfo {author} {\bibfnamefont {J.}~\bibnamefont
  {Cortes}}, \ and\ \bibinfo {author} {\bibfnamefont {R.}~\bibnamefont
  {Vilela~Mendes}},\ }\href {\doibase 10.1103/PhysRevE.102.013210} {\bibfield
  {journal} {\bibinfo  {journal} {Phys. Rev. E}\ }\textbf {\bibinfo {volume}
  {102}},\ \bibinfo {pages} {013210} (\bibinfo {year} {2020})}\BibitemShut
  {NoStop}%
\bibitem [{\citenamefont {Chaves}\ \emph {et~al.}(2017)\citenamefont {Chaves},
  \citenamefont {Peres}, \citenamefont {Smirnov},\ and\ \citenamefont
  {Mortensen}}]{graf6}%
  \BibitemOpen
  \bibfield  {author} {\bibinfo {author} {\bibfnamefont {A.~J.}\ \bibnamefont
  {Chaves}}, \bibinfo {author} {\bibfnamefont {N.~M.~R.}\ \bibnamefont
  {Peres}}, \bibinfo {author} {\bibfnamefont {G.}~\bibnamefont {Smirnov}}, \
  and\ \bibinfo {author} {\bibfnamefont {N.~A.}\ \bibnamefont {Mortensen}},\
  }\href {\doibase 10.1103/physrevb.96.195438} {\bibfield  {journal} {\bibinfo
  {journal} {Physical Review B}\ }\textbf {\bibinfo {volume} {96}} (\bibinfo
  {year} {2017}),\ 10.1103/physrevb.96.195438}\BibitemShut {NoStop}%
\bibitem [{\citenamefont {Atwal}\ and\ \citenamefont
  {Ashcroft}(2002)}]{conserved}%
  \BibitemOpen
  \bibfield  {author} {\bibinfo {author} {\bibfnamefont {G.~S.}\ \bibnamefont
  {Atwal}}\ and\ \bibinfo {author} {\bibfnamefont {N.~W.}\ \bibnamefont
  {Ashcroft}},\ }\href {\doibase 10.1103/PhysRevB.65.115109} {\bibfield
  {journal} {\bibinfo  {journal} {Phys. Rev. B}\ }\textbf {\bibinfo {volume}
  {65}},\ \bibinfo {pages} {115109} (\bibinfo {year} {2002})}\BibitemShut
  {NoStop}%
\bibitem [{\citenamefont {Svintsov}\ \emph {et~al.}(2013)\citenamefont
  {Svintsov}, \citenamefont {Vyurkov}, \citenamefont {Ryzhii},\ and\
  \citenamefont {Otsuji}}]{graf7}%
  \BibitemOpen
  \bibfield  {author} {\bibinfo {author} {\bibfnamefont {D.}~\bibnamefont
  {Svintsov}}, \bibinfo {author} {\bibfnamefont {V.}~\bibnamefont {Vyurkov}},
  \bibinfo {author} {\bibfnamefont {V.}~\bibnamefont {Ryzhii}}, \ and\ \bibinfo
  {author} {\bibfnamefont {T.}~\bibnamefont {Otsuji}},\ }\href {\doibase
  10.1103/physrevb.88.245444} {\bibfield  {journal} {\bibinfo  {journal}
  {Physical Review B}\ }\textbf {\bibinfo {volume} {88}} (\bibinfo {year}
  {2013}),\ 10.1103/physrevb.88.245444}\BibitemShut {NoStop}%
\bibitem [{\citenamefont {Svintsov}(2018)}]{rapidc}%
  \BibitemOpen
  \bibfield  {author} {\bibinfo {author} {\bibfnamefont {D.}~\bibnamefont
  {Svintsov}},\ }\href {\doibase 10.1103/PhysRevB.97.121405} {\bibfield
  {journal} {\bibinfo  {journal} {Phys. Rev. B}\ }\textbf {\bibinfo {volume}
  {97}},\ \bibinfo {pages} {121405} (\bibinfo {year} {2018})}\BibitemShut
  {NoStop}%
\bibitem [{cgr()}]{cgraf7}%
  \BibitemOpen
  \href@noop {} {\bibinfo  {journal} {In fact, referring to Ref.~\cite{graf7}
  and following its Authors, their Eq.~$(15)$ is obtained by dividing
  Eqs.~$(5)$ and $(9)$ therein, which prescription results not in said Eq.
  (15), but rather in our Eq.~\eqref{masssmall}}\ }\BibitemShut {NoStop}%
\bibitem [{\citenamefont {Lee}\ and\ \citenamefont {Jung}(2017)}]{bohmfinal}%
  \BibitemOpen
\bibfield  {journal} {  }\bibfield  {author} {\bibinfo {author} {\bibfnamefont
  {M.-J.}\ \bibnamefont {Lee}}\ and\ \bibinfo {author} {\bibfnamefont {Y.-D.}\
  \bibnamefont {Jung}},\ }\href {\doibase
  https://doi.org/10.1016/j.physleta.2016.12.025} {\bibfield  {journal}
  {\bibinfo  {journal} {Physics Letters A}\ }\textbf {\bibinfo {volume}
  {381}},\ \bibinfo {pages} {636 } (\bibinfo {year} {2017})}\BibitemShut
  {NoStop}%
\bibitem [{\citenamefont {Akbari-Moghanjoughi}(2012)}]{DensePlasmas1}%
  \BibitemOpen
  \bibfield  {author} {\bibinfo {author} {\bibfnamefont {M.}~\bibnamefont
  {Akbari-Moghanjoughi}},\ }\href {\doibase 10.1063/1.3699535} {\bibfield
  {journal} {\bibinfo  {journal} {Physics of Plasmas}\ }\textbf {\bibinfo
  {volume} {19}},\ \bibinfo {pages} {042701} (\bibinfo {year} {2012})},\
  \Eprint {http://arxiv.org/abs/https://doi.org/10.1063/1.3699535}
  {https://doi.org/10.1063/1.3699535} \BibitemShut {NoStop}%
\bibitem [{\citenamefont {Manfredi}\ and\ \citenamefont {Haas}(2001)}]{hydro1}%
  \BibitemOpen
  \bibfield  {author} {\bibinfo {author} {\bibfnamefont {G.}~\bibnamefont
  {Manfredi}}\ and\ \bibinfo {author} {\bibfnamefont {F.}~\bibnamefont
  {Haas}},\ }\href {\doibase 10.1103/PhysRevB.64.075316} {\bibfield  {journal}
  {\bibinfo  {journal} {Phys. Rev. B}\ }\textbf {\bibinfo {volume} {64}},\
  \bibinfo {pages} {075316} (\bibinfo {year} {2001})}\BibitemShut {NoStop}%
\bibitem [{\citenamefont {Li}(2010)}]{DensePlasmas2}%
  \BibitemOpen
  \bibfield  {author} {\bibinfo {author} {\bibfnamefont {S.-C.}\ \bibnamefont
  {Li}},\ }\href {\doibase 10.1063/1.3476275} {\bibfield  {journal} {\bibinfo
  {journal} {Physics of Plasmas}\ }\textbf {\bibinfo {volume} {17}},\ \bibinfo
  {pages} {082307} (\bibinfo {year} {2010})}\BibitemShut {NoStop}%
\bibitem [{\citenamefont {Nicholson}(1983)}]{nicholson_1983}%
  \BibitemOpen
  \bibfield  {author} {\bibinfo {author} {\bibfnamefont {D.}~\bibnamefont
  {Nicholson}},\ }\href {https://books.google.pt/books?id=fyRRAAAAMAAJ} {\emph
  {\bibinfo {title} {Introduction to Plasma Theory}}}\ (\bibinfo  {publisher}
  {Wiley},\ \bibinfo {year} {1983})\BibitemShut {NoStop}%
\bibitem [{\citenamefont {Avron}(1998)}]{avron_1998}%
  \BibitemOpen
  \bibfield  {author} {\bibinfo {author} {\bibfnamefont {J.~E.}\ \bibnamefont
  {Avron}},\ }\href {\doibase 10.1023/a:1023084404080} {\bibfield  {journal}
  {\bibinfo  {journal} {Journal of Statistical Physics}\ }\textbf {\bibinfo
  {volume} {92}},\ \bibinfo {pages} {543} (\bibinfo {year} {1998})}\BibitemShut
  {NoStop}%
\bibitem [{\citenamefont {Souslov}\ \emph {et~al.}(2019)\citenamefont
  {Souslov}, \citenamefont {Dasbiswas}, \citenamefont {Fruchart}, \citenamefont
  {Vaikuntanathan},\ and\ \citenamefont {Vitelli}}]{souslov}%
  \BibitemOpen
  \bibfield  {author} {\bibinfo {author} {\bibfnamefont {A.}~\bibnamefont
  {Souslov}}, \bibinfo {author} {\bibfnamefont {K.}~\bibnamefont {Dasbiswas}},
  \bibinfo {author} {\bibfnamefont {M.}~\bibnamefont {Fruchart}}, \bibinfo
  {author} {\bibfnamefont {S.}~\bibnamefont {Vaikuntanathan}}, \ and\ \bibinfo
  {author} {\bibfnamefont {V.}~\bibnamefont {Vitelli}},\ }\href {\doibase
  10.1103/PhysRevLett.122.128001} {\bibfield  {journal} {\bibinfo  {journal}
  {Phys. Rev. Lett.}\ }\textbf {\bibinfo {volume} {122}},\ \bibinfo {pages}
  {128001} (\bibinfo {year} {2019})}\BibitemShut {NoStop}%
\end{thebibliography}%

\end{document}